\documentclass[12pt]{article}
\usepackage[utf8]{inputenc}
\usepackage{amsfonts}
\usepackage{amsmath}
\usepackage{amssymb}
\usepackage{graphicx}
\usepackage{cite}
\usepackage{color}
\usepackage[left=1cm,top=1cm,bottom=1cm,right=1cm,nohead,nofoot]{geometry}

\def \be {\begin{equation}}
\def \ee {\end{equation}}
\def \bea {\begin{eqnarray}}
\def \eea {\end{eqnarray}}
\def \nn {\nonumber}

\def \rr {\raise.35ex\hbox{\small $\prime$}\kern-.17em{\mbox{\large $\imath$}}}
\def \del {\partial}
\def \dels {\partial\kern-.6em /\kern.1em}
\def \As {{A\kern-.5em / \kern.5em}}
\def \Ds {D\kern-.7em / \kern.5em}
\def \a {\alpha}

\def \b {\beta}

\def \eps {\epsilon}

\def \ks {k\kern-.5em /}
\def \ls {l\kern-.5em /}\def \lam {\lambda}

\def \dm {\dot{\mu}}

\def \mn {\mu\nu}
\def \del {\partial}

\reversemarginpar
\newcommand{\hide}[1]{}

\begin{document}
\begin{titlepage}

\begin{center}

\hfill
\vskip .2in

\textbf{\LARGE
Electric-Magnetic Dualities in Non-Abelian and Non-Commutative Gauge Theories
\vskip.5cm
}

\vskip .5in
{\large
Jun-Kai Ho$^a$ \footnote{e-mail address: junkai125@gmail.com} and
Chen-Te Ma$^b$ \footnote{e-mail address: yefgst@gmail.com}\\
\vskip 3mm
}
{\sl
${}^a$
Department of Physics, Brown University, Box 1843 Providence, RI 02912-1843, USA,\\
${}^b$
Department of Physics and Center for Theoretical Sciences, 
National Taiwan University, Taipei 10617, Taiwan, R.O.C.
}\\
\vskip 3mm
\vspace{60pt}
\end{center}
\begin{abstract}
Electric-magnetic dualities are equivalence between strong and weak coupling constants. A standard example is the exchange of electric and magnetic fields in an abelian gauge theory. We show three methods to perform electric-magnetic dualities in the case of the non-commutative $U(1)$ gauge theory. The first method is to use covariant field strengths to be the electric and magnetic fields. We find an invariant form of an equation of motion after performing the electric-magnetic duality. The second method is to use the Seiberg-Witten map to rewrite the non-commutative $U(1)$ gauge theory in terms of abelian field strength. The third method is to use the large Neveu Schwarz-Neveu Schwarz (NS-NS) background limit (non-commutativity parameter only has one degree of freedom) to consider the non-commutative $U(1)$ gauge theory or D3-brane. In this limit, we introduce or dualize a new one-form gauge potential to get a D3-brane in a large Ramond-Ramond (R-R) background via field redefinition. We also use perturbation to study the equivalence between two D3-brane theories. Comparison of these methods in the non-commutative $U(1)$ gauge theory gives different physical implications. The comparison reflects the differences between the non-abelian and non-commutative gauge theories in the electric-magnetic dualities. For a complete study, we also extend our studies to the simplest abelian and non-abelian $p$-form gauge theories, and a non-commutative theory with the non-abelian structure. 

\end{abstract}

\end{titlepage}

\section{Introduction}
\label{1}
The M-theory provides useful dualities to unify different kinds of theories and helps us to understand supergravity solutions \cite{Gauntlett:2003cy}. In low-energy limit, the ten dimensional supergravity has the T-duality and S-duality. The T-duality is a duality on a target space. The T-duality of closed string theory \cite{Zwiebach:1992ie} exchanges the momentum and winding modes, and the T-duality of open string theory exchanges the Dirichlet and Neumann boundary conditions. The T-duality requires an isometry on a compact torus, but a generic background does not always have an isometry in closed string theory. In other words, the T-duality maps single valued fields to non-single valued fields and we lose periodicity of the background. Then we obtain the non-geometric flux after performing the T-duality twice in the case of constant $H$-flux. This mapping gives rises to a problem on quantum dynamics. The solution is to use a double space to construct a well-defined transition function as a diffeomorphism in closed string theory \cite{Hull:2009mi}. With a global symmetry description, we sacrifice local symmetry in the double space. Local symmetry in the double space is still possible, but difficulties come from the closure of the generalized Lie derivative. This double construction is also extended to open string theory, and has also been applied to cosmology \cite{Ma:2014ala, Ma:2014vqm, Ma:2015yma}. These formulations rely on geometric constructions from the Courant bracket or generalized geometry \cite{Gualtieri:2003dx}. This bracket comes from the combination of tangent and cotangent bundles. A theory in a double space with the strong constraints (removing additional coordinates) is equivalent to a theory with the Courant bracket. The S-duality is a non-perturbative duality by exchanging the strong and weak coupling constants. In four dimensional electromagnetism, we have an electric-magnetic duality between electric and magnetic fields. This duality is a special case of the S-duality. A problem with the S-duality is that it is hard to be performed exactly due to some issues involving strong couplings. At low-energy level, one successful example is a low-energy effective theory with a non-commutativity parameter (inversely proportional to antisymmetric backgrounds) being a perturbative parameter \cite{Ho:2013opa}. The extension of duality from ten dimensional supergravity to eleven dimensional supergravity is the U-duality combining T-duality and S-duality. The manifest U-duality is studied in \cite{Berman:2010is} using extended coordinates.

String theory is described by a two dimensional sigma model. On bulk, the sigma model describes gravity. When we impose the Dirichlet and Neumann boundary conditions on the sigma model, the boundary term comes from the gauge principle. This boundary term gives a picture of open string ending on a D-brane. The ending point of the open string shows the non-commutativity. Non-commutative geometry is naturally hidden in string theory. The low-energy effective theory \cite{Ho:2013opa, Ho:2008nn, Ho:2013paa, Ho:2014una, Zwiebach:1985uq} of open string is the Dirac-Born-Infeld (DBI) model. In the DBI model, we have the Seiberg-Witten map that maps the commutative theory to the non-commutative theory. In the non-commutative description, the leading order term in the action is a non-commutative $U(1)$ gauge theory with the Moyal product. The Moyal product captures all the effects of the non-commutativity parameters. We find an alternative way to examine the string theory. Now we have many different kinds of non-commutative geometry generalized from the DBI model. This generalization helps us to find more interesting field theories and constrain our low-energy effective field theories from the non-commutative geometry. The first example is the Nambu-Poisson M5 (NP M5) brane theory. This theory describes a M2-M5 system in the large C field background (only three spatial components) on the non-commutative space at low-energy level \cite{Ho:2008nn}. Based on dimensional reduction, we find a D$p$-brane in the large ($p$-1)-form background \cite{Ho:2013paa} and a D$p$-brane in the large NS-NS two-form background. Especially for $p=3$, the S-duality relation to all orders is found in \cite{Ho:2013opa}. According to the dualities, we find the S-duality relation and the non-commutative geometry on the R-R background. The second example is the non-commutative geometry in closed string theory. The Seiberg-Witten map and the Moyal product in the DBI model rely on one-form gauge transformation. A low-energy effective theory of the double sigma model shows a combination of two-form antisymmetric background field and two-form field strength on boundary and bulk \cite{Ma:2014vqm, Ma:2015yma}. We have the one-form gauge transformation on the bulk in the low-energy effective theory without using the strong constraints. This shows a non-trivial existence of the Seiberg-Witten map and Moyal product on the bulk. The non-commutative geometry in open string theory can easily describe all background effects from the Moyal product in the non-commutative descriptions. We should obtain all $\alpha^{\prime}$ effects from the Moyal product or the non-commutative geometry. 

A low-energy effective theory of open string at leading order is the abelian Yang-Mills theory. The abelian Yang-Mills theory in four dimensions at classical level describes the famous Maxwell's equations. This theory has local gauge symmetry, and its equation of motion is gauge invariant. An extension of a gauge principle from the abelian gauge group to the non-abelian gauge group gives the non-abelian Yang-Mills theory. An ordinary derivative operator in the abelian Yang-Mills theory becomes a covariant derivative operator in the non-abelian Yang-Mills theory. The gauge invariant property of the field strength and equation of motion are modified accordingly. The non-abelian Yang-Mills theory has a gauge covariant field strength and a corresponding equation of motion. The gauge principle also helps us to find open string. Local gauge symmetry has a very long history in aiding the construction of new theories and simplifying our analysis. But local gauge symmetry has its own loophole due to redundant descriptions. This situation implies that local gauge symmetry is too restricted. We never observe gauge symmetry in our nature. The observed fact is that photon has two polarization states. Violating the local gauge symmetry is not equivalent to violating our experimental results. An interesting symmetry constraint should contain physical information and should not be too restrictive to kill off interactions. Global symmetry is a good candidate. When we gauge fix a theory, the gauge fixing term does not break the global symmetry. The global symmetry gives more structures and the Noether currents to our theories. The Noether currents are important ingredients for the conserved quantities. Double field theory combines diffeomorphism and one-form gauge transformation to form an $O(D, D)$ global structure in a double space. This is an example to define the T-duality in a generic background from global symmetry to avoid isometry problem. Electric-magnetic duality for the abelian group in four dimensions only exchanges electric and magnetic fields. This is a rotation-like symmetry so electric-magnetic duality should be the global symmetry in the abelian gauge theories. Global symmetry is a physical symmetry, so a full study of electric-magnetic dualities should be interesting. 

We use three methods to study electric-magnetic dualities in the non-commutative $U(1)$ gauge theory. The first way is to use covariant field strength as the electric and magnetic fields. The second method \cite{Ganor:2000my} is to use the Seiberg-Witten map to change variables in terms of the abelian field strength. This result is interesting because the non-commutative $U(1)$ gauge theory has a non-abelian-like structure which comes from the Moyal product. This structure should forbid us to perform the electric-magnetic duality. The Seiberg-Witten map helps us to rewrite the non-commutative $U(1)$ gauge theory in a suitable form to perform the electric-magnetic duality. This method sheds light on finding some hidden symmetry structures to understand the electric-magnetic dualities in the non-abelian gauge theories. The third method is to consider the large NS-NS and R-R background limit. In these limits, a D3-brane in the large NS-NS background is equivalent to a D3-brane in the large R-R background under the electric-magnetic duality. We use field redefinition and perturbation to check the electric-magnetic duality in this method. Although they give different physical interpretations in these methods, they are all interesting to find mappings between a strongly and weakly coupled gauge theories. The non-commutative $U(1)$ gauge theory is a good toy model to study electric-magnetic dualities. Although this theory does not have a non-abelian gauge group, the Moyal product produces a non-abelian-like term. The electric-magnetic dualities are very different between the abelian and non-abelian gauge theories. Equations of motion do not depend on gauge potentials in abelian gauge theories, but equations of motion in non-abelian gauge theories do. A standard electric-magnetic duality is to exchange the electric and magnetic fields. If equations of motion depend on gauge potentials, the standard electric-magnetic duality should not work. A direct generalization should exchange the gauge potentials to find a dual action at quantum level \cite{GarciaCompean:1998nt}. We can also put the gauge and dual gauge fields together to find the manifest electric-magnetic duality in an abelian gauge theory \cite{Ma:2015yma}. This direct generalization is our first method. This method can be performed in the non-commutative $U(1)$ gauge and non-abelian Yang-Mills theories. The electric and magnetic fields in the non-abelian gauge and non-commutative $U(1)$ gauge theories are covariant objects. They are not gauge invariant as abelian gauge theories. In abelian gauge theories, electric and magnetic fields are physical observables. A magnetic monopole solution in the abelian Yang-Mills theory should be detectable if magnetic monopoles exist in our nature. But the magnetic monopole solution for field strength in the non-abelian Yang-Mills theory is not a detectable observable.  In our first method, we can find more differences between abelian and non-abelian gauge theories. In abelian gauge theories, we have a restriction on dimensionality from the Poincaré lemma. But we do not have the Poincaré lemma in non-abelian gauge theories. We lose a restriction on dimensionality. This feature possibly reflects the fact that the electric-magnetic dualities have different interpretations in interacting theories. The second and third methods are also suitable in the non-commutative $U(1)$ gauge theory. A good property of these methods is that we have a restriction on dimensionality for the non-commutative $U(1)$ gauge theory. But they cannot be extended to the non-abelian gauge theories. The second and third methods imply that the non-abelian-like term in the non-commutative $U(1)$ gauge theory is still different from the non-abelian term in the non-abelian Yang-Mills theory. We compactify 2-torus in the multiple M5-branes theory, then we should obtain two D3-branes with different backgrounds arising from the ordering of compactification. There is S-duality or electric-magnetic duality between two theories. More suitable and consistent electric-magnetic dualities should help us to probe a consistent multiple M5-branes theory. We will point out the difficulty in our studies. For a generic study and completeness, we also define the electric-magnetic dualities in the simplest $p$-form gauge theory with the abelian and non-abelian gauge groups, and a non-commutative theory with the non-abelian structures. 

We first review the electric-magnetic duality of the abelian and non-abelian Yang-Mills theories in Sec. \ref{2}. Then we give three ways to perform the electric-magnetic dualities of the non-commutative $U(1)$ gauge theory in Sec. \ref{3}. The extension of the electric-magnetic duality of the $p$-form gauge theory with abelian and non-abelian groups, and a non-commutative theory with a non-abelian structure are in Sec. \ref{4}. Finally, we conclude and discuss in Sec. \ref{5}.


\section{Review of the Electric-Magnetic Duality in the Abelian and Non-Abelian Yang-Mills Theories}
\label{2}
We review the electric-magnetic dualities for the abelian and non-abelian Yang-Mills theories \cite{GarciaCompean:1998nt} in this section. The electric-magnetic dualities in the abelian and non-abelian Yang-Mills theories exchange the gauge and dual gauge fields. The gauge field in the equations of motion is simply replaced by the dual gauge field under electric-magnetic duality. A difference between the two theories is a restriction on dimensionality from the Poincaré lemma. This restriction only exists in the abelian gauge theory. Since the non-abelian structure contains an interaction term, the Poincaré lemma is no longer valid to constrain dimensionality. In the non-abelian Yang-Mills theory, this approach has one advantage that the electric and magnetic fields are the covariant field strengths.
\subsection{Abelian Yang-Mills Theory}
The abelian Yang-Mills theory is
\begin{equation}
S_{\mbox{AB}}=-\frac{1}{4g_{YM}^2}\int d^4 x~ F_{\mu\nu} F^{\mu\nu},
\end{equation}
where $F_{\mu\nu}=\del_\mu A_\nu-\del_\nu A_\mu$, and $g_{YM}$ is gauge coupling constant. We denote spacetime indices by the Greek letters. 

 We introduce an antisymmetric auxiliary field, $G_{\mn}$, this action is written as 
\begin{align}
& \int d^4 x~ \left(g_{YM}^2G_{\mn} G^{\mn}-G^{\mn} F_{\mn}\right) \nn\\
=&\int d^4 x ~\bigg(g_{YM}^2G_{\mn} G^{\mn}-G^{\mn}(\del_\mu A_\nu-\del_\nu A_\mu)\bigg).
\end{align}
Then we integrate out $A$ to obtain
\bea
\int{\cal  D}G\exp\bigg\lbrack ig_{YM}^2\int d^4 x ~\bigg(G_{\mn} G^{\mn}\bigg)\bigg\rbrack\delta\bigg(\partial_{\mu}G^{\mu\nu}\bigg).
\eea
Because of
\bea
\partial_{\mu}G^{\mu\nu}=\frac{1}{2}\epsilon^{\mu\nu\rho\sigma}\partial_{\mu}\tilde{G}_{\rho\sigma}=0,
\eea
we get
\bea
\tilde{G}_{\mu\nu}=\partial_{\mu}\tilde{A}_{\nu}-\partial_{\nu}\tilde{A}_{\mu}
\eea
from the Poincaré lemma. Solving the delta function, we obtain
\bea
-\frac{g_{YM}^2}{4}\int d^4 x~ \tilde{G}_{\mu\nu} \tilde{G}^{\mu\nu}.
\eea
At classical level, we find
\bea
\partial_{\mu}F^{\mu\nu}=0~\longleftrightarrow~\partial_{\mu}\tilde{G}^{\mu\nu}=0.
\eea
This is the familiar electric-magnetic duality without source.
We use the Poincaré lemma to obtain the restriction on dimensionality. However, we have another method to perform the electric-magnetic duality for the abelian Yang-Mills theory in all dimensions. We start from
\bea
\int{\cal  D}G\exp\bigg\lbrack ig_{YM}^2\int d^4 x ~\bigg(G_{\mn} G^{\mn}\bigg)\bigg\rbrack\delta\bigg(\partial_{\mu}G^{\mu\nu}\bigg).
\eea
Then we introduce an auxiliary field $\tilde{A}$ to rewrite the partition function as
\bea
&&\int{\cal  D}G{\cal D} \tilde{A}\exp\bigg\lbrack ig_{YM}^2\int d^4 x ~\bigg(G_{\mn} G^{\mn}-2\partial_{\mu}\tilde{A}_{\nu}G^{\mu\nu}\bigg)\bigg\rbrack
\nn\\
&=&
\int{\cal  D}G{\cal D} \tilde{A}\exp\bigg\lbrack ig_{YM}^2\int d^4 x ~\bigg(G_{\mn} G^{\mn}-\big(\partial_{\mu}\tilde{A}_{\nu}-\partial_{\nu}\tilde{A}_{\mu}\big)G^{\mu\nu}\bigg)
\bigg\rbrack.
\eea
Hence, we integrate $G_{\mu\nu}$ out to obtain
\bea
-\frac{g_{YM}^2}{4}\int d^4 x~ \tilde{G}_{\mu\nu} \tilde{G}^{\mu\nu}.
\eea
Because we do not use the Poincaré lemma to perform the electric-magnetic duality, we can use this method to extend the electric-magnetic duality from four to all dimensions. We will also apply this method to the non-abelian Yang-Mills gauge theory.

\subsection{Non-Abelian Yang-Mills Theory}
The action for the Non-abelian Yang-Mills theory is
\begin{equation}
S_{\mbox{NAB}}=-\frac{1}{4g_{YM}^2}\int d^4 x~ F_{\mu\nu}^{a} F^{\mu\nu,a},
\end{equation}
where $F_{\mu\nu}^a=\del_\mu A_\nu^a-\del_\nu A_\mu^a +\lbrack A_{\mu}, A_{\nu}\rbrack^a$. We define $\lbrack A_{\mu}, A_{\nu}\rbrack^a\equiv f^{abc}A_\mu^b A_\nu^c$, and  denote the Lie algebra indices from $a$ to $z$. By introducing an antisymmetric auxiliary field $G_{\mn}^a$, this action can be written as 
\begin{align}
&\int d^4 x~ \left(g_{YM}^2G_{\mn}^a G^{\mn,a}-G^{\mn,a} F_{\mn}^a\right) \nn\\
=&\int d^4 x ~\bigg\lbrack g_{YM}^2G_{\mn}^a G^{\mn,a}-G^{\mn,a}\bigg(\del_\mu A_\nu^a-\del_\nu A_\mu^a +f^{abc}A_\mu^b A_\nu^c\bigg)\bigg\rbrack\nn\\
=&\int d^4 x ~\bigg(g_{YM}^2G_{\mn}^a G^{\mn,a}+2G^{\mn,a}\del_\nu A_\mu^a -A_\mu^b f^{abc} G^{\mn,a} A_\nu^c\bigg)\nn\\
=&\int d^4 x ~\bigg(g_{YM}^2G_{\mn}^a G^{\mn,a}-2\del_\nu G^{\mn,a} A_\mu^a -A_\mu^b f^{abc} G^{\mn,a} A_\nu^c \bigg).
\end{align}
The action is quadratic in $A$, then we can integrate it out  in path integral by using the Gaussian integral 
\begin{equation}
\label{Gaussian}
\int {\cal D}x~ e^{\frac{i}{2}x^T M x +i Jx}\sim \sqrt{\frac{1}{\det M}} e^{-\frac{i}{2}J^T M^{-1} J}.
\end{equation}
The partition function becomes
\begin{equation}
\label{nonabelian_Z1}
Z\sim \int \mathcal{D}G~ \big(\det M\big)^{-\frac{1}{2}} \exp \bigg\lbrack ig_{YM}^2\int d^4 x~\bigg(G_{\mn}^a G^{\mn,a}+ \del_\gamma G^{\mu\gamma,a} (M^{-1})_{\mn}^{ab} \del_\lambda G^{\nu\lambda,b} \bigg)\bigg\rbrack,
\nn\\
\end{equation}
where $M^{\mn,bc}=g_{YM}^2 f^{abc} G^{\mn,a}$.
Let us define $\bar{A}_\mu^a\equiv-(M^{-1})_{\mn}^{ab} \del_\rho G^{\nu\rho,b}$. Therefore, $G^{\mn}$ satisfies an equation of motion
\begin{align}
\del_\nu G^{\nu\mu,b}+M^{\mn,ab}\bar{A}_\nu^a =\del_\nu G^{\nu\mu,b}+ g_{YM}^2 f^{cab} G^{\mn,c} \bar{A}_\nu^a=\del_\nu G^{\nu\mu,b}- g_{YM}^2 f^{abc}\bar{A}_\nu^a G^{\nu\mu,c} &=0.
\end{align}

Now, we consider 
\begin{align*}
\int d^4x~G^{\mn,a}F_{\mn}^a(\bar{A})&=\int d^4x~G^{\mn,a}\big(\del_\mu \bar{A}_\nu^a-\del_\nu \bar{A}_\mu^a + f^{\prime abc}\bar{A}_\mu^b \bar{A}_\nu^c\big)\nn\\
&=\int d^4x~\bigg\lbrack-2 G^{\mn,a}\del_\nu \bar{A}_\mu^a +f^{\prime abc} G^{\mn,a} \bar{A}_\mu^b \bar{A}_\nu^c\bigg\rbrack \nn\\
&=\int d^4x~ \bigg(2\del_\nu G^{\mn,a} \bar{A}_\mu^a + f^{\prime abc} G^{\mn,a} \bar{A}_\mu^b \bar{A}_\nu^c\bigg) \nn\\
&=\int d^4x~\bigg\lbrack-2 \del_\nu G^{\mn,a} (M^{-1})_{\mu\lambda}^{ab} \del_\rho G^{\lambda\rho,b}+M^{\mn,bc} (M^{-1})_{\mu\lambda}^{bd} \del_\rho G^{\lambda\rho,d} (M^{-1})^{ce}_{\nu\gamma} \del_\rho G^{\gamma\rho,e}\bigg\rbrack \nn\\
&=\int d^4 x~\bigg(- \del_\nu G^{\mn,a} (M^{-1})_{\mu\lambda}^{ab} \del_\rho G^{\lambda\rho,b}\bigg),
\end{align*}
where $f^{\prime abc}=g^2_{YM}f^{abc}$.
Therefore, the partition function becomes
\begin{equation}
\label{nonabelian_Z2}
Z\sim\int \mathcal{D}G \mathcal{D}\bar{A}~ (\det M)^{-\frac{1}{2}} \exp \bigg\lbrack ig_{YM}^2\int d^4 x~\bigg(G_{\mn}^a G^{\mn,a}- G^{\mn,a}F_{\mn}^a\big(\bar{A}\big) \bigg) \bigg\rbrack\delta\bigg(2\bar{A}_\mu^a+2(M^{-1})_{\mn}^{ab} \del_\rho G^{\nu\rho,b}\bigg),
\end{equation}
where the factor 2 in the delta function is introduced for convenience.
We can write the delta function in exponential form. For the convenience of integration, we can write the delta function in the other way as
\begin{equation}
\delta(2\bar{A}+2M^{-1}\del G)=\delta\bigg(M^{-1}(2M\bar{A}+2\del G)\bigg).
\end{equation}
This extracts a factor of $\det M$ out of the delta function. Then we get
\begin{equation}
\label{nonabelian_Z3}
Z\sim \int \mathcal{D}G \mathcal{D}\bar{A}\mathcal{D}\Lambda~ (\det M)^{\frac{1}{2}}\exp \bigg\lbrack ig_{YM}^2\int d^4 x~\bigg(G_{\mn}^a G^{\mn,a}- G^{\mn,a}F_{\mn}^a(\bar{A})+2\Lambda_\mu^a (M^{\mn,ab}\bar{A}_\nu^b+\del_\rho G^{\mu\rho,a}) \bigg) \bigg\rbrack.
\end{equation}
The last bracket in the exponential can be simplified as
\begin{align}
2\int d^4x~\bigg(\Lambda_\mu^a (M^{\mn,ab}\bar{A}_\nu^b+\del_\rho G^{\mu\rho,a})\bigg)&=2\int d^4x~\bigg(\Lambda_\mu^a M^{\mn,ab}\bar{A}_\nu^b-\del_\rho \Lambda_\mu^aG^{\mu\rho,a}\bigg)\nn\\
&=2\int d^4x~\bigg(\Lambda_\mu^a g_{YM}^2 f^{cab}G^{\mn,c} \bar{A}_\nu^b-\del_\rho \Lambda_\mu^aG^{\mu\rho,a}\bigg) \nn\\
&=2\int d^4x~\bigg\lbrack-G^{\mn,a}\bigg(\del_\nu \Lambda_\mu^a-g_{YM}^2 f^{bac}\bar{A}_\nu^b \Lambda_\mu^c\bigg)\bigg\rbrack\nn\\
&=-2\int d^4x~\bigg(  G^{\mn,a} \big(D_\nu^{(\bar{A})}\Lambda_\mu\big)^a\bigg) \nn\\
&=2\int d^4x~\bigg(G^{\mn,a}\big( D_\mu^{(\bar{A})}\Lambda_\nu\big)^a\bigg).
\end{align}
Substitution of this term into \eqref{nonabelian_Z3} gives
\begin{equation}
Z\sim\int \mathcal{D}G ~(\det M)^{\frac{1}{2}} \int \mathcal{D}\bar{A}\mathcal{D}\Lambda~\exp \Bigg\lbrack ig_{YM}^2\int d^4 x~\bigg\lbrack G^{\mn,a}\bigg(G_{\mn,a}- F_{\mn}^a(\bar{A})+2\big(D_\mu^{(\bar{A})}\Lambda_\nu\big)^a\bigg) \bigg\rbrack \Bigg\rbrack.
\end{equation}
Let us change the variable to $\tilde{A}_\mu=\bar{A}_\mu-\Lambda_\mu$. The field strength can be written as
\begin{align}
F_{\mn}(\bar{A})=&\del_\mu(\tilde{A}_\nu^a+\Lambda_\nu^a)-\del_\nu(\tilde{A}_\mu^a+\Lambda_\mu^a)+f^{\prime abc}(\tilde{A}_\mu^b+\Lambda_\mu^b)(\tilde{A}_\nu^c+\Lambda_\nu^c)\nn\\
=&\del_\mu \tilde{A}_\nu^a-\del_\nu \tilde{A}_\mu^a+f^{\prime abc}\tilde{A}_\mu^b \tilde{A}_\nu^c\nn\\
&+\del_\mu \Lambda_\nu^a-\del_\nu \Lambda_\mu^a+2 f^{\prime abc}\tilde{A}_\mu^b \Lambda_\nu^c+f^{\prime abc}\tilde{A}_\nu^c \Lambda_\mu^b+ f^{\prime abc}\Lambda_\mu^b \Lambda_\nu^c \nn\\
=&F_{\mn}^a(\tilde{A})+D_\mu^{(\bar{A})} \Lambda_\nu^a- D_\nu^{(\bar{A})} \Lambda_\mu^a+f^{\prime abc}\Lambda_\mu^b \Lambda_\nu^c. 
\end{align}
Thus, we obtain
\begin{align}
Z\sim &\int \mathcal{D}G ~(\det M)^{\frac{1}{2}} \int \mathcal{D}\tilde{A}\mathcal{D}\Lambda~\exp \Bigg\lbrack ig_{YM}^2\int d^4 x~\bigg\lbrack G^{\mn,a}\bigg(G_{\mn}^a- F_{\mn}^a(\tilde{A})-f^{\prime abc}\Lambda_\mu^b \Lambda_\nu^c \bigg) \bigg\rbrack \Bigg\rbrack\nn\\
=&\int \mathcal{D}G ~(\det M)^{\frac{1}{2}} \int \mathcal{D}\tilde{A}~\exp \bigg\lbrack ig_{YM}^2\int d^4 x~ G^{\mn,a}\bigg(G_{\mn}^a- F_{\mn}^a(\tilde{A})\bigg) \bigg\rbrack 
\nn\\
&\times\int \mathcal{D}\Lambda ~\exp \bigg(- i g_{YM}^2 G^{\mn,a}f^{\prime abc}\Lambda_\mu^b \Lambda_\nu^c  \bigg)
\nn\\
=&\int \mathcal{D}G ~(\det M)^{\frac{1}{2}} \int \mathcal{D}\tilde{A}~\exp \Bigg( ig_{YM}^2\int d^4 x~\bigg\lbrack G^{\mn,a}\bigg(G_{\mn}^a- F_{\mn}^a(\tilde{A})\bigg)\bigg\rbrack \Bigg)
\nn\\
&\times \int \mathcal{D}\Lambda ~\exp \bigg(- ig_{YM}^2 \Lambda_\mu^a M^{\mn,ab} \Lambda_\nu^b \bigg)
\nn\\
\sim &\int \mathcal{D}G \mathcal{D}\tilde{A}~\exp \bigg\lbrack ig_{YM}^2\int d^4 x~ G^{\mn,a}\bigg(G_{\mn}^a- F_{\mn}^a(\tilde{A})\bigg) \bigg\rbrack .
\end{align}
To get the last line, we integrate the field $\Lambda$ out and get a factor of $(\det M)^{-1/2}$, which cancels $(\det M)^{1/2}$. This result shows the covariance of the partition function by comparing the partition functions. We equivalently obtain 
\bea
D_{\mu}^{(A)}F^{\mu\nu}(A)=0~\longleftrightarrow~D_{\mu}^{(\tilde{A})}F^{\mu\nu}(\tilde{A})=0
\eea
at classical level. In the non-abelian Yang-Mills theory, the equation of motion depends on the gauge potential. The abelian Yang-Mills theory only relies on the field strength at classical level. In the abelian Yang-Mills theory, we can use the the Poincaré lemma to perform the electric-magnetic duality at classical level. But we cannot do in the non-abelian Yang-Mills theory because the equation of motion is related to a gauge potential. This shows that the electric-magnetic duality is more delicate in the non-abelian Yang-Mills theory than in the abelian Yang-Mills theory. Although we consider four dimensions in the case of the non-abelian Yang-Mills theory, we can extend from four dimensions to arbitrary dimensions. Since we do not have the Poincaré lemma at the non-abelian level, there is no constraint on the number of dimensions. In this method, we use covariant field strength to be the electric and magnetic fields. The covariant quantities are not physical quantities. This property points out one difference of the electric-magnetic duality between the non-abelian and abelian gauge theories. 

\section{Electric-Magnetic Dualities in the Non-Commutative $U(1)$ Gauge Theory}
\label{3}
We use three methods to perform the electric-magnetic dualities for the non-commutative $U(1)$ gauge theory. The first approach is to use the covariant field strength to be the electric and magnetic fields. Then we will obtain a similar answer like in the non-abelian Yang-Mills theory. The non-commutative $U(1)$ gauge theory has a non-abelian-like structure which comes from the Moyal product so we should obtain a similar answer for the electric-magnetic duality. The second method is to implement the electric-magnetic duality by the Seiberg-Witten map. This map transforms a non-commutative theory to a commutative theory. In the third method, we consider the large background limit to perform the electric-magnetic duality from field redefinition and perturbation. In these three methods, we can observe that the non-commutative $U(1)$ gauge theory is different from the non-abelian Yang-Mills theory because the second and third methods cannot be applied to the non-abelian Yang-Mills theory. 

\subsection{The First Method}
The action for the non-commutative $U(1)$ gauge theory is
\begin{equation}
S_{\mbox{NC}}=-\frac{1}{4g_{YM}^2}\int d^4 x~ \hat{F}_{\mu\nu}*\hat{F}^{\mu\nu},
\end{equation}
where $\hat{F}_{\mu\nu}=\del_\mu \hat{A}_\nu-\del_\nu \hat{A}_\mu +\lbrack \hat{A}_\mu,\hat{A}_\nu\rbrack_*$ is a non-commutative field strength, $\hat{A}$ is the non-commutative gauge potential, and $*$ is the star product. The star product is defined by
\bea
A*B&\equiv& A\exp\bigg(\frac{\theta^{\mu\nu}}{2} \overleftarrow{\partial}_{\mu}\overrightarrow{\partial}_{\nu}\bigg)B,
\nn\\
\lbrack A, B\rbrack_*&\equiv& A*B-B*A,
\eea
where $\theta^{\mu\nu}$ is a constant non-commutativity parameter. In string theory, the non-commutativity parameter is inversely proportional to a B-field background if the B-field background is large.

By introducing an antisymmetric auxiliary field $\hat{G}_{\mn}$, this action can be rewritten as 
\begin{equation}
\label{action}
S=\int d^4 x~ \left(g_{YM}^2\hat{G}_{\mn}*\hat{G}^{\mn}-\hat{G}^{\mn} *\hat{F}_{\mn}\right).
\end{equation}
Using a formula
\begin{equation}
\int d^4 x~ f*g=\int d^4 x~ fg,
\end{equation}
the action becomes
\begin{align}
S&=\int d^4 x~ \left(g_{YM}^2 \hat{G}_{\mn}\hat{G}^{\mn}-\hat{G}^{\mn}\hat{F}_{\mn}\right) \nn\\
&=\int d^4 x ~\bigg\lbrack g_{YM}^2\hat{G}_{\mn}\hat{G}^{\mn}-\hat{G}^{\mn}\bigg(\del_\mu \hat{A}_\nu-\del_\nu \hat{A}_\mu +\lbrack\hat{A}_\mu,\hat{A}_\nu\rbrack_*\bigg)\bigg\rbrack\nn\\
&\approx \int d^4 x ~\Bigg\lbrack g_{YM}^2\hat{G}_{\mn}\hat{G}^{\mn}-\hat{G}^{\mn}\bigg\lbrack\del_\mu \hat{A}_\nu-\del_\nu \hat{A}_\mu +\frac{1}{2}\theta^{\rho\sigma}\bigg(\del_\rho \hat{A}_\mu \del_\sigma \hat{A}_\nu-\del_\rho \hat{A}_\nu \del_\sigma \hat{A}_\mu\bigg)\bigg\rbrack\Bigg\rbrack\nn\\
&=\int d^4 x ~\left(g_{YM}^2\hat{G}_{\mn}\hat{G}^{\mn}+2\hat{G}^{\mn}\del_\nu \hat{A}_\mu -\theta^{\rho\sigma}\hat{G}^{\mn} \del_\rho \hat{A}_\mu \del_\sigma \hat{A}_\nu\right)\nn\\
&=\int d^4 x ~\left(g_{YM}^2\hat{G}_{\mn}\hat{G}^{\mn}-2\del_\nu\hat{G}^{\mn} \hat{A}_\mu +\theta^{\rho\sigma} \hat{A}_\mu \del_\rho \hat{G}^{\mn}  \del_\sigma \hat{A}_\nu+ \theta^{\rho\sigma} \hat{A}_\mu \hat{G}^{\mn} \del_\rho  \del_\sigma \hat{A}_\nu \right)\nn\\
&=\int d^4 x ~\left(g_{YM}^2\hat{G}_{\mn}\hat{G}^{\mn}-2\del_\nu\hat{G}^{\mn} \hat{A}_\mu +\theta^{\rho\sigma} \hat{A}_\mu \del_\rho \hat{G}^{\mn}  \del_\sigma \hat{A}_\nu \right),
\end{align}
where we just consider the action up to the first order of $\theta$ and ignore total derivative terms. We used antisymmetric property of $G_{\mn}$ to get the fourth line from the third line. We integrate by part from the fourth line to the fifth line. The last term in the fifth line vanishes because of the antisymmetric property of $\theta^{\rho\sigma}$. Now, the action is quadratic in the field $A$, then we can integrate this field out in path integral by using the Gaussian integral \eqref{Gaussian}.

The partition function is given by
\begin{equation}
\label{Z1_1}
Z\sim \int \mathcal{D}G~ (\det M)^{-\frac{1}{2}} \exp \bigg\lbrack ig_{YM}^2\int d^4 x~\bigg( \hat{G}_{\mn}\hat{G}^{\mn}- \del_\gamma \hat{G}^{\mu\gamma} \big(M^{-1}\big)_{\mn} \del_\lambda \hat{G}^{\nu\lambda} \bigg)\bigg\rbrack,
\end{equation}
where $M^{\mn}=g_{YM}^2\theta^{\rho\sigma}\del_\rho \hat{G}^{\mn}\del_\sigma$.
Let us define $\bar{A}_\mu\equiv(M^{-1})_{\mn} \del_\rho \hat{G}^{\nu\rho}$, from which this turns out that $\hat{G}^{\mn}$ satisfies the equation of motion up to the first order (in the Poisson limit),
\bea
&&\del_\nu \hat{G}^{\nu\mu}+M^{\mn}\bar{A}_\nu=0 \nn\\
&\Rightarrow& \del_\nu \hat{G}^{\nu\mu}+g_{YM}^2 \theta^{\rho\sigma}\del_\rho \hat{G}^{\mn}\del_\sigma \bar{A}_\nu=0 \nn\\
&\Rightarrow& \del_\nu \hat{G}^{\nu\mu}+ \{\bar{A}_\nu,\hat{G}^{\nu\mu}\}=0,
\eea
where $\{A, B\}\equiv g_{YM}^2 \theta^{\mu\nu}\partial_{\mu}A\partial_{\nu}B\equiv\tilde{\theta}^{\mu\nu}\partial_{\mu}A\partial_{\nu}B$.

Let us consider this term
\begin{align*}
\int d^4x~\hat{G}^{\mn}\hat{F}_{\mn}(\bar{A})&=\int d^4x~\hat{G}^{\mn}\bigg(\del_\mu \bar{A}_\nu-\del_\nu \bar{A}_\mu + \{\bar{A}_\mu,\bar{A}_\nu\}\bigg)\nn\\
&=\int d^4x~\bigg(-2 \hat{G}^{\mn}\del_\nu \bar{A}_\mu + \hat{G}^{\mn} \tilde{\theta}^{\rho\sigma}\del_\rho \bar{A}_\mu \del_\sigma \bar{A}_\nu\bigg) \nn\\
&=\int d^4x~\bigg(2\del_\nu \hat{G}^{\mn} \bar{A}_\mu -\bar{A}_\mu \tilde{\theta}^{\rho\sigma} \del_\rho \hat{G}^{\mn}\del_\sigma \bar{A}_\nu\bigg) \nn\\
&=\int d^4x~\bigg(2 \del_\nu \hat{G}^{\mn} \big(M^{-1}\big)_{\mu\lambda} \del_\rho \hat{G}^{\lambda\rho}-\big(M^{-1}\big)_{\mu\lambda} \del_\rho \hat{G}^{\lambda\rho} (M)^{\mn}\big(M^{-1}\big)_{\nu\sigma}\del_\delta \hat{G}^{\sigma\delta}\bigg) \nn\\
&=\int d^4x~\del_\nu \hat{G}^{\mn} (M^{-1})_{\mu\lambda} \del_\rho \hat{G}^{\lambda\rho},
\end{align*}
where we used integration by part from the second to the third line and substituted $\bar{A}_\mu=(M^{-1})_{\mn} \del_\rho \hat{G}^{\nu\rho}$ into the fourth line. This term is equal to the second term in the partition function \eqref{Z1_1}.
Therefore, the partition function can be rewritten as
\begin{equation}
\label{Z2}
Z\sim\int \mathcal{D}G ~(\det M)^{-\frac{1}{2}} \int \mathcal{D}\bar{A}~\exp \left( ig_{YM}^2\int d^4 x~\Big( \hat{G}_{\mn}\hat{G}^{\mn}- \hat{G}^{\mn}\hat{F}_{\mn}(\bar{A}) \Big) \right)\delta\bigg(2\bar{A}_\rho-2(M^{-1})_{\rho\sigma} \del_\lambda \hat{G}^{\sigma\lambda}\bigg),
\end{equation}
where the factor of 2 in the delta function is introduced for convenience. We express the delta function as
\begin{equation}
\delta(2\bar{A}-2M^{-1}\del \hat{G})=\delta\Big(M^{-1}(2M\bar{A}-2\del \hat{G})\Big).
\end{equation}
This extracts a factor $\det M$ out of the delta function after integrating. Therefore, we get
\begin{equation}
\label{Z3}
Z\sim \int \mathcal{D}G ~(\det M)^{\frac{1}{2}} \int \mathcal{D}\bar{A}\mathcal{D}\Lambda~\exp \Bigg\lbrack ig_{YM}^2\int d^4 x~\bigg\lbrack \hat{G}_{\mn}\hat{G}^{\mn}- \hat{G}^{\mn}\hat{F}_{\mn}(\bar{A})-\Lambda_\mu \bigg(2M^{\mn}\bar{A}_\nu-2\del_\rho \hat{G}^{\mu\rho}\bigg) \bigg\rbrack \Bigg\rbrack.
\end{equation}
The last bracket in the exponential can be simplified as
\begin{align}
\int d^4x~2\bigg\lbrack\Lambda_\mu\bigg(M^{\mn}\bar{A}_\nu-\del_\rho\hat{G}^{\mu\rho}\bigg)\bigg\rbrack&=\int d^4x~\bigg(2\Lambda_\mu M^{\mn}\bar{A}_\nu+2\del_\rho \Lambda_\mu\hat{G}^{\mu\rho}\bigg)\nn\\
&=\int d^4x~\bigg(2\Lambda_\mu \tilde{\theta}^{\rho\sigma}\del_\rho \hat{G}^{\mn}\del_\sigma \bar{A}_\nu+2\del_\rho \Lambda_\mu\hat{G}^{\mu\rho}\bigg) \nn\\
&=\int d^4x~\bigg(- 2\hat{G}^{\mn} \tilde{\theta}^{\rho\sigma} \del_\rho \Lambda_\mu \del_\sigma \bar{A}_\nu+2\del_\rho \Lambda_\mu\hat{G}^{\mu\rho}\bigg)\nn\\
&=\int d^4x~\bigg(2\hat{G}^{\mn} \times \{\bar{A}_\nu,\Lambda_\mu\} +2(\del_\nu \Lambda_\mu)\hat{G}^{\mu\nu}\bigg)\nn\\
&=\int d^4x~\bigg(2\hat{G}^{\mn} D_\nu^{(\bar{A})} \Lambda_\mu\bigg) \nn\\
&=\int d^4x~\bigg(-2\hat{G}^{\mn} D_\mu^{(\bar{A})} \Lambda_\nu\bigg),
\end{align}
where we define $D_\mu^{(\bar{A})}O\equiv\del_\mu O+\{\bar{A}_\mu, O\}$. Substitution of this term into the partition function gives
\begin{equation}
Z\sim\int \mathcal{D}G ~(\det M)^{1/2} \int \mathcal{D}\bar{A}\mathcal{D}\Lambda~\exp \Bigg\lbrack ig_{YM}^2\int d^4 x~\bigg\lbrack \hat{G}^{\mn}\bigg(\hat{G}_{\mn}- \hat{F}_{\mn}(\bar{A})+2D_\mu^{(\bar{A})} \Lambda_\nu \bigg) \bigg\rbrack \Bigg\rbrack.
\end{equation}
Let us define a new variable $\tilde{A}_\mu\equiv\bar{A}_\mu-\Lambda_\mu$. The field strength can be written as
\begin{align}
\hat{F}_{\mn}(\bar{A})=&\del_\mu(\tilde{A}_\nu+\Lambda_\nu)-\del_\nu(\tilde{A}_\mu+\Lambda_\mu)+\{\tilde{A}_\mu+\Lambda_\mu,\tilde{A}_\nu+\Lambda_\nu\}\nn\\
=&\del_\mu \tilde{A}_\nu-\del_\nu \tilde{A}_\mu+\{\tilde{A}_\mu,\tilde{A}_\nu\}\nn\\
&+\del_\mu \Lambda_\nu-\del_\nu \Lambda_\mu +\{\tilde{A}_\mu,\Lambda_\nu\}+\{\Lambda_\mu,\tilde{A}_\nu\}+\{\Lambda_\mu,\Lambda_\nu\} \nn\\
=&\hat{F}_{\mn}(\tilde{A})+D_\mu^{(\bar{A})} \Lambda_\nu-D_\nu^{(\bar{A})} \Lambda_\mu +\tilde{\theta}^{\rho\sigma}\del_\rho \Lambda_\mu \del_\sigma \Lambda_\nu.
\end{align}
Thus, we get
\begin{align}
Z\sim &\int \mathcal{D}G ~(\det M)^{\frac{1}{2}} \int \mathcal{D}\tilde{A}\mathcal{D}\Lambda~\exp \Bigg\lbrack ig_{YM}^2\int d^4 x~\bigg\lbrack \hat{G}^{\mn}\bigg(\hat{G}_{\mn}- \hat{F}_{\mn}\big(\tilde{A}\big)-\tilde{\theta}^{\rho\sigma}\del_\rho \Lambda_\mu \del_\sigma \Lambda_\nu\bigg) \bigg\rbrack \Bigg\rbrack\nn\\
=&\int \mathcal{D}G ~(\det M)^{\frac{1}{2}} \int \mathcal{D}\tilde{A}~\exp \Bigg\lbrack ig_{YM}^2\int d^4 x~\bigg\lbrack \hat{G}^{\mn}\bigg(\hat{G}_{\mn}- \hat{F}_{\mn}\big(\tilde{A}\big)\bigg)\bigg\rbrack \Bigg\rbrack 
\nn\\
&\times\int \mathcal{D}\Lambda ~\exp \bigg\lbrack -i\int d^4x~\bigg(g_{YM}^2\hat{G}^{\mn}\tilde{\theta}^{\rho\sigma}\del_\rho \Lambda_\mu \del_\sigma \Lambda_\nu\bigg) \bigg\rbrack\nn\\
=&\int \mathcal{D}G ~(\det M)^{\frac{1}{2}} \int \mathcal{D}\tilde{A}~\exp \Bigg\lbrack ig_{YM}^2\int d^4 x~\bigg\lbrack \hat{G}^{\mn}\bigg(\hat{G}_{\mn}- \hat{F}_{\mn}\big(\tilde{A}\big)\bigg) \bigg\rbrack\Bigg\rbrack
\nn\\
&\times\int \mathcal{D}\Lambda ~\exp \bigg\lbrack i\int d^4x~ \bigg(g_{YM}^2 \Lambda_\mu \tilde{\theta}^{\rho\sigma}\del_\rho \hat{G}^{\mn}  \del_\sigma \Lambda_\nu\bigg) \bigg\rbrack\nn\\
=&\int \mathcal{D}G (\det M)^{\frac{1}{2}} \int \mathcal{D}\tilde{A}~\exp \Bigg\lbrack ig_{YM}^2\int d^4 x~\bigg\lbrack \hat{G}^{\mn}\bigg(\hat{G}_{\mn}- \hat{F}_{\mn}\big(\tilde{A}\big)\bigg)\bigg\rbrack \Bigg\rbrack 
\nn\\
&\times\int \mathcal{D}\Lambda \exp \bigg(ig_{YM}^2 \int d^4x~ \Lambda_\mu M^{\mn} \Lambda_\nu \bigg)\nn\\
\sim &\int \mathcal{D}G \mathcal{D}\tilde{A}~\exp \Bigg\lbrack ig_{YM}^2\int d^4 x~\bigg\lbrack \hat{G}^{\mn}\bigg(\hat{G}_{\mn}- \hat{F}_{\mn}\big(\tilde{A}\big)\bigg)\bigg\rbrack \Bigg\rbrack .
\end{align}
To get the last line, we integrate the field $\Lambda$ out and obtain a factor of $(\det M)^{-1/2}$ which cancels the factor $(\det M)^{1/2}$ in front of the measure. This calculation shows 
\bea
D_{\mu}^{A}\hat{F}^{\mu\nu}(A)=0~\longleftrightarrow~D_{\mu}^{\tilde{A}}\hat{F}^{\mu\nu}(\tilde{A})=0
\eea
at classical level.
We can also extend the result from four dimensions to all dimensions as in the case of the non-abelian Yang-Mills theory because we do not use any information related to the Poincaré lemma. The non-commutative $U(1)$ gauge theory has a non-abelian-like structure, which comes from the Moyal product so it is not surprising to obtain a similar answer from this method. We will show two other methods to perform the electric-magnetic dualities in the non-commutative $U(1)$ gauge theory. We will eventually find that these two methods cannot be applied to the non-abelian Yang-Mills theory.
\subsection{The Second Method}
An equation of motion in the non-commutative $U(1)$ theory depends on a gauge potential. This property causes some difficulties to define the electric-magnetic dualities. From a point of view of string theory, the non-commutative geometry can be connected to the commutative geometry via the Seiberg-Witten map. We can use this Seiberg-Witten map to redefine our theory in terms of abelian field strength on the commutative space. The Seiberg-Witten map is defined on a commutative diagram. We first use gauge transformation, then redefine (Seiberg-Witten map) the theory from the commutative to the non-commutative gauge fields. On the other hand, we change the ordering. We first redefine the theory from the commutative gauge to non-commutative gauge fields, then we perform the gauge transformation on the non-commutative space. These operations should be equivalent  because gauge transformation and redefinition do not change any physical meaning. Then we can find a condition for the Seiberg-Witten map as
\bea
\hat{A}(A)+\hat{\delta}_{\hat{\lambda}}(A)=\hat{A}(A+\delta_{\lambda}A),
\eea  
where $\hat{A}$ is the Seiberg-Witten map, $\delta_{\lambda}$ is a gauge transformation on the commutative space and $\hat{\delta}_{\hat{\lambda}}$ is a gauge transformation on the non-commutative space. Let us define the gauge transformations
\bea
\delta_{\lambda}A_{\mu}\equiv\partial_{\mu}\lambda, \qquad \hat{\delta}_{\hat{\lambda}}\hat{A}_{\mu}\equiv\partial_{\mu}\hat{\lambda}-\lbrack\hat{\lambda}, \hat{A}_{\mu}\rbrack_*.
\eea
Now we calculate $\hat{A}$ and $\hat{\lambda}$ at leading order. For convenience, we define
\bea
\hat{A}\equiv A+A^{\prime}(A), \qquad \hat{\lambda}\equiv\lambda+\lambda^{\prime},
\eea
where $A^{\prime}$ and $\lambda^{\prime}$ are higher-order effects. If we consider first order correction with respect to $\theta$, the condition for the Seiberg-Witten map becomes
\bea
\label{SW}
A^{\prime}_{\mu}(A+\delta_{\lambda}A)-A^{\prime}_{\mu}(A)-\partial_{\mu}\lambda^{\prime}=-\theta^{\rho\sigma}\partial_{\rho}\lambda\partial_{\sigma}A_{\mu}.
\eea
We find a solution
\bea
\hat{A}_{\mu}=A_{\mu}-\theta^{\rho\sigma}\bigg(A_{\rho}\partial_{\sigma}A_{\mu}-\frac{1}{2}A_{\rho}\partial_{\mu}A_{\sigma}\bigg), \qquad \hat{\lambda}=\lambda+\frac{1}{2}\theta^{\rho\sigma}A_{\sigma}\partial_{\rho}\lambda .
\eea
From this solution, we get
\bea
A^{\prime}_{\mu}(A+\delta_{\lambda}A)&=&-\theta^{\rho\sigma}\bigg\lbrack\bigg(A_{\rho}+\partial_{\rho}\lambda\bigg)\partial_{\sigma}\bigg(A_{\mu}+\partial_{\mu}\lambda\bigg)-\frac{1}{2}\bigg(A_{\rho}+\partial_{\rho}\lambda\bigg)\partial_{\mu}\bigg(A_{\sigma}+\partial_{\sigma}\lambda\bigg)\bigg\rbrack,
\nn\\
A^{\prime}_{\mu}(A)&=&-\theta^{\rho\sigma}\bigg(A_{\rho}\partial_{\sigma}A_{\mu}-\frac{1}{2}A_{\rho}\partial_{\mu}A_{\sigma}\bigg),
\nn\\
\partial_{\mu}\lambda^{\prime}&=&\frac{1}{2}\theta^{\rho\sigma}\big(\partial_{\mu}\partial_{\rho}\lambda\big) A_{\sigma}+\frac{1}{2}\theta^{\rho\sigma}\partial_{\rho}\lambda\partial_{\mu}A_{\sigma}.
\eea 
We can check this solution by plugging these terms into the left hand side of (\ref{SW}) and considering the first order in $\lambda$ to obtain
\bea
-\theta^{\rho\sigma}\partial_{\rho}\lambda\partial_{\sigma}A_{\mu}.
\eea
Now we use this solution to consider $\hat{F}$ in the Poisson limit as
\bea
\hat{F}_{\mu\nu}&\approx& \partial_{\mu}\hat{A}_{\nu}-\partial_{\nu}\hat{A}_{\mu}+\theta^{\rho\sigma}\partial_{\rho}\hat{A}_{\mu}\partial_{\sigma}\hat{A}_{\nu}
\nn\\
&\approx&\partial_{\mu}A_{\nu}-\theta^{\rho\sigma}\bigg(\partial_{\mu}A_{\rho}\partial_{\sigma}A_{\nu}+A_{\rho}\partial_{\mu}\partial_{\sigma}A_{\nu}-\frac{1}{2}\partial_{\mu}A_{\rho}\partial_{\nu}A_{\sigma}-\frac{1}{2}A_{\rho}\partial_{\mu}\partial_{\nu}A_{\sigma}\bigg)
\nn\\
&&
-\partial_{\nu}A_{\mu}+\theta^{\rho\sigma}\bigg(\partial_{\nu}A_{\rho}\partial_{\sigma}A_{\mu}+A_{\rho}\partial_{\nu}\partial_{\sigma}A_{\mu}-\frac{1}{2}\partial_{\nu}A_{\rho}\partial_{\mu}A_{\sigma}-\frac{1}{2}A_{\rho}\partial_{\nu}\partial_{\mu}A_{\sigma}\bigg)
\nn\\
&&
+\theta^{\rho\sigma}\partial_{\rho}A_{\mu}\partial_{\sigma}A_{\nu}
\nn\\
&=&
\partial_{\mu}A_{\nu}-\partial_{\nu}A_{\mu}-\theta^{\rho\sigma}\bigg(A_{\rho}\partial_{\sigma}F_{\mu\nu}+\partial_{\mu}A_{\rho}\partial_{\sigma}A_{\nu}
-\partial_{\mu}A_{\rho}\partial_{\nu}A_{\sigma}-\partial_{\rho}A_{\mu}\partial_{\sigma}A_{\nu}-\partial_{\nu}A_{\rho}\partial_{\sigma}A_{\mu}\bigg)
\nn\\
&=&F_{\mu\nu}+\theta^{\rho\sigma}\bigg(F_{\mu\rho}F_{\nu\sigma}-A_{\rho}\partial_{\sigma}F_{\mu\nu}\bigg).
\eea
We find one solution from the Poisson limit to infinite orders as
\bea
\delta\hat{A}_{\mu}&=&-\frac{1}{4}\delta\theta^{\rho\sigma}\bigg\lbrack\hat{A}_{\rho}*\bigg(2\partial_{\sigma}\hat{A}_{\mu}-\partial_{\mu}\hat{A}_{\sigma}\bigg)
+\bigg(2\partial_{\sigma}\hat{A}_{\mu}-\partial_{\mu}\hat{A}_{\sigma}\bigg)*\hat{A}_{\rho}\bigg\rbrack,
\nn\\
\delta\hat{\lambda}&=&\frac{1}{4}\delta\theta^{\rho\sigma}\bigg(\partial_{\rho}\hat{\lambda}*\hat{A}_{\sigma}+\hat{A}_{\rho}*\partial_{\sigma}\hat{\lambda}\bigg),
\nn\\
\delta\hat{F}_{\mu\nu}&=&\frac{1}{4}\delta\theta^{\rho\sigma}\bigg\lbrack2\hat{F}_{\mu\rho}*\hat{F}_{\nu\sigma}+2\hat{F}_{\nu\sigma}*\hat{F}_{\mu\rho}
-\hat{A}_{\rho}*\bigg(\partial_{\sigma}\hat{F}_{\mu\nu}+\hat{D}_{\sigma}\hat{F}_{\mu\nu}\bigg)-\bigg(\partial_{\sigma}\hat{F}_{\mu\nu}
+\hat{D}_{\sigma}\hat{F}_{\mu\nu}\bigg)*\hat{A}_{\rho}\bigg\rbrack,
\nn\\
\eea
where 
\bea
\hat{D}_{\lambda}\hat{F}_{\mu\nu}\equiv\partial_{\lambda}\hat{F}_{\mu\nu}+\lbrack\hat{A}_{\lambda}, \hat{F}_{\mu\nu}\rbrack_*.
\eea
Then we use this Seiberg-Witten map to change variables to write the theory in terms of abelian field strength as
\bea
-\frac{1}{4g_{YM}^2}\int d^4x~\hat{F}_{\mu\nu}\hat{F}^{\mu\nu}\approx\frac{1}{g_{YM}^2}\int d^4x~\bigg(-\frac{1}{4}F^{\mu\nu}F_{\mu\nu}+\frac{1}{2}F^{\mu\nu}F_{\mu\rho}\theta^{\rho\sigma}F_{\sigma\nu}+\frac{1}{2}F^{\mu\nu}A_{\rho}\theta^{\rho\sigma}\partial_{\sigma}F_{\mu\nu}\bigg).
\nn\\
\eea
If we ignore total derivative terms, the final term can be rewritten as
\bea
\frac{1}{2}\int d^4x~F^{\mu\nu}A_{\rho}\theta^{\rho\sigma}\partial_{\sigma}F_{\mu\nu}&=&\int d^4x~\bigg(-\frac{1}{2}\partial_{\sigma}F^{\mu\nu}A_{\rho}\theta^{\rho\sigma}F_{\mu\nu}-\frac{1}{2}F^{\mu\nu}\partial_{\sigma}A_{\rho}\theta^{\rho\sigma}F_{\mu\nu}\bigg)
\nn\\
&=&\int d^4x~\bigg(-\frac{1}{4}F^{\mu\nu}F_{\sigma\rho}\theta^{\rho\sigma}F_{\mu\nu}-\frac{1}{2}\partial_{\sigma}F^{\mu\nu}A_{\rho}\theta^{\rho\sigma}F_{\mu\nu}\bigg)
\nn\\
&=&\int d^4x~\bigg(\frac{1}{4}\mbox{tr}(\theta F)\mbox{tr}(F^2)-\frac{1}{2}\partial_{\sigma}F^{\mu\nu}A_{\rho}\theta^{\rho\sigma}F_{\mu\nu}\bigg).
\eea
Hence, we get
\bea
\int d^4x~F^{\mu\nu}A_{\rho}\theta^{\rho\sigma}\partial_{\sigma}F_{\mu\nu}=\frac{1}{4}\int d^4x~\mbox{tr}(\theta F)\mbox{tr}(F^2).
\eea
Therefore, the action is
\bea
-\frac{1}{4g_{YM}^2}\int d^4x~\hat{F}_{\mu\nu}\hat{F}^{\mu\nu}\approx-\frac{1}{4g_{YM}^2}\int d^4x~\bigg(F^{\mu\nu}F_{\mu\nu}+2\mbox{tr}(\theta F^3)-\frac{1}{2}\mbox{tr}(\theta F)\mbox{tr}(F^2)\bigg).
\nn\\
\eea
Then we add one additional term to change $F$ to be an unconstrained field as
\bea
\frac{1}{g_{YM}^2}\int d^4x~\bigg(-\frac{1}{4}F^{\mu\nu}F_{\mu\nu}+\frac{1}{2}F^{\mu\nu}F_{\mu\rho}\theta^{\rho\sigma}F_{\sigma\nu}-\frac{1}{8}F^{\mu\nu}F_{\sigma\rho}\theta^{\rho\sigma}F_{\mu\nu}+\frac{1}{2}\tilde{G}_{\mu\nu}F^{\mu\nu}\bigg),
\nn\\
\eea
where $\tilde{G}_{\mu\nu}\equiv\frac{1}{2}\epsilon_{\mu\nu\rho\sigma}G^{\rho\sigma}$.
If we integrate $G$ out, we can obtain $dF=0$. Then we solve $dF=0$ to go back to the original action.  Varying $F$ gets an equation of motion for $F$ as
\bea
&&-F_{\mu\nu}+F_{\mu\rho}\theta^{\rho\sigma}F_{\sigma\nu}+F_{\mu\rho}\theta_{\nu\sigma}F^{\sigma\rho}+F_{\sigma\nu}F^{\sigma\rho}\theta_{\rho\mu}-\frac{1}{2}F_{\sigma\rho}\theta^{\rho\sigma}F_{\mu\nu}-\frac{1}{4}F^{\rho\sigma}\theta_{\mu\nu}F_{\rho\sigma}+g_{YM}^2\tilde{G}_{\mu\nu}=0
\nn\\
&\rightarrow&F_{\mu\nu}=g_{YM}^2\tilde{G}_{\mu\nu}-F_{\mu\rho}\theta^{\rho\sigma}F_{\sigma\nu}-F_{\mu\rho}\theta_{\nu\sigma}F^{\sigma\rho}-F_{\sigma\nu}F^{\sigma\rho}\theta_{\rho\mu}+\frac{1}{2}F_{\sigma\rho}\theta^{\rho\sigma}F_{\mu\nu}+\frac{1}{4}F^{\rho\sigma}\theta_{\mu\nu}F_{\rho\sigma}.
\nn\\
\eea
If we only consider first order with respect to $\theta$, the action is
\bea
\int d^4x~\bigg(\frac{g_{YM}^2}{4}\tilde{G}_{\mu\nu}\tilde{G}^{\mu\nu}+\frac{g_{YM}^4}{2}\tilde{G}^{\mu\nu}\tilde{G}_{\mu\rho}\theta^{\rho\sigma}\tilde{G}_{\sigma\nu}-\frac{g_{YM}^4}{8}\tilde{G}^{\mu\nu}\tilde{G}_{\sigma\rho}\theta^{\rho\sigma}\tilde{G}_{\mu\nu}\bigg).
\eea
The first term can be written as
\bea
\frac{g_{YM}^2}{4}\int d^4x~\tilde{G}_{\mu\nu}\tilde{G}^{\mu\nu}=-\frac{g_{YM}^2}{4}\int d^4x~G_{\mu\nu}G^{\mu\nu}.
\eea
Then we define $g_{YM}^2\theta^{\rho\sigma}\equiv -\frac{1}{2}\epsilon^{\rho\sigma\rho^{\prime\prime}\sigma^{\prime\prime}}\tilde{\theta}_{\rho^{\prime\prime}\sigma^{\prime\prime}}$ to rewrite the second and third terms in the action. The second term is
\bea
\frac{g_{YM}^4}{2}\int d^4x~\tilde{G}^{\mu\nu}\tilde{G}_{\mu\rho}\theta^{\rho\sigma}\tilde{G}_{\sigma\nu}
&=&-\frac{g_{YM}^2}{8}\int d^4x~\tilde{G}^{\mu\nu}\tilde{G}_{\mu\rho}\tilde{\theta}_{\rho^{\prime\prime}\sigma^{\prime\prime}}G^{\sigma^{\prime}\nu^{\prime}}
\epsilon^{\sigma\rho\rho^{\prime\prime}\sigma^{\prime\prime}}\epsilon_{\sigma\nu\sigma^{\prime}\nu^{\prime}}
\nn\\
&=&g_{YM}^2\int d^4x~\bigg(\frac{1}{4}\tilde{G}^{\mu\nu}\tilde{G}_{\mu\nu}\tilde{\theta}_{\rho^{\prime\prime}\sigma^{\prime\prime}}
G^{\rho^{\prime\prime}\sigma^{\prime\prime}}+\frac{1}{2}\tilde{G}^{\mu\nu}\tilde{G}_{\mu\rho}\tilde{\theta}_{\nu\sigma^{\prime}}G^{\sigma^{\prime}\rho}\bigg)
\nn\\
&=&g_{YM}^2\int d^4x~\bigg(-\frac{1}{4}G^{\mu\nu}G_{\mu\nu}\tilde{\theta}_{\rho\sigma}G^{\rho\sigma}-\frac{1}{4}G^{\mu\nu}G_{\mu\nu}\tilde{\theta}_{\rho\sigma}G^{\sigma\rho}
\nn\\
&&+\frac{1}{2}G^{\mu^{\prime}\rho}G_{\mu^{\prime}\nu}\tilde{\theta}^{\nu\sigma^{\prime}}G_{\sigma^{\prime}\rho}\bigg)
\nn\\
&=&-\frac{g_{YM}^2}{2}\int d^4x~\mbox{tr}\big(\tilde{\theta}G^3\big).
\eea
The third term can be rewritten as
\bea
-\frac{g_{YM}^4}{8}\int d^4x~\tilde{G}^{\mu\nu}\tilde{G}_{\sigma\rho}\theta^{\rho\sigma}\tilde{G}_{\mu\nu}&=&\frac{g_{YM}^4}{16}\int d^4x~G^{\mu\nu}G_{\mu\nu}\epsilon_{\sigma\rho\sigma^{\prime}\rho{\prime}}\theta^{\rho\sigma}G^{\sigma^{\prime}\rho^{\prime}}=-\frac{g_{YM}^2}{8}\int d^4x~G^{\mu\nu}G_{\mu\nu}\tilde{\theta}_{\sigma^{\prime}\rho^{\prime}}G^{\sigma^{\prime}\rho^{\prime}}
\nn\\
&=&-\frac{g_{YM}^2}{8}\int d^4x~\mbox{tr}\big(G^2\big)\mbox{tr}\big(\tilde{\theta}G\big).
\eea
Therefore, the action is given by
\bea
g_{YM}^2\int d^4x~\bigg(-\frac{1}{4}G_{\mu\nu}G^{\mu\nu}-\frac{1}{2}\mbox{tr}\big(\tilde{\theta}G^3\big)-\frac{1}{8}\mbox{tr}\big(G^2\big)\mbox{tr}\big(\tilde{\theta}G\big)\bigg)\approx -\frac{g_{YM}^2}{4}\int d^4x~\hat{G}_{\mu\nu}\hat{G}^{\mu\nu}
\nn\\
\eea
after we perform the electric-magnetic duality. The result is very simple and interesting. The electric-magnetic duality just exchanges $\theta$ and $\tilde{\theta}$ and invert the gauge coupling constant. Here, $\tilde{\theta}$ is not the same as in the first method. The reason possibly comes from the use of the Poincaré lemma. The lemma gives a standard dual between electric and magnetic fields. In the second method, we use the Seiberg-Witten map to rewrite the non-commutative theory in terms of the abelian field strength to let us exchange electric and magnetic fields directly. But the first method loses some information from the Poincaré lemma, then we map all ordinary field strengths to all dual field strengths. This is why we can use the first method to perform the electric-magnetic duality in all dimensions, but the second method is only valid in four dimensions.

The use of the Seiberg-Witten map can rewrite the non-commutative $U(1)$ gauge theory in terms of the abelian field strength. This is amazing and surprising. An equation of motion in the non-commutative $U(1)$ gauge theory depends on the gauge potential. Naively, we should encounter some difficulties as in the non-abelian gauge theories. But we can use the Seiberg-Witten map to rewrite our theories in terms of the abelian field strength to avoid these problems in the non-commutative $U(1)$ gauge theory. This possibly reminds us that we can perform some operations in the non-abelian gauge theories from some hidden symmetry structures. However, we cannot use the Seiberg-Witten map in the non-commutative $U(N)$ gauge theory to perform the same electric-magnetic duality. Although the non-commutative $U(1)$ gauge theory has a non-abelian-like structure, the non-abelian-like structure comes from the derivative operation. This non-abelian-like structure is still different from the non-abelian structure. If we can find a way to relate the gauge potential via field redefinition in the non-commutative $U(1)$ gauge theory, we might apply this method to the non-abelian gauge theories. We will introduce this method in the next section.
\subsection{The Third Method}
We consider the D3-brane in the large NS-NS two-form background. This theory on the non-commutative space is described by the non-commutative $U(1)$ gauge theory. A well-known fact is that we can perform the S-duality or electric-magnetic duality to get the D3-brane in the large R-R two-form background. These two theories in the Poisson limit come from different orderings of compactified directions in the Nambu-Poisson M5-brane. Different orderings of compactified directions should not change physical meaning. We first perform a field redefinition from the NS-NS D3-brane theory to R-R D3-brane theory \cite{Ho:2013opa}. Then we perform an electric-magnetic duality from the R-R D3-brane theory to the NS-NS D3 brane theory to show their equivalence.

\subsubsection{Field Redefinition}
In this section, we use field redefinition to connect two D3-brane theories \cite{Ho:2013opa}, which come from different orderings of compactified directions in the Nambu-Poisson M5-brane theory. The Nambu-Poisson M5-brane theory is a well-defined theory under the large C-field background. After we compactify 2-torus, the D3-brane should be well-defined under the large NS-NS two-form background or large R-R two-form background. From a string point of view, we should use the electric-magnetic duality to connect them. In other words, we can have a field redefinition to connect them. Due to this large background, we have two kinds of spacetime directions in our theories. Our conventions of world-volume indices are $\a, \b = 0, 1$, $\dot\mu, \dot\nu = \dot1, \dot2$ 
and $A, B = 0, 1, \dot1, \dot2$. The dotted indices denote the directions that 
the NS-NS B-field (or R-R field) background is turned on. The large two-form background only opens on two spatial directions. Other components of the two-form background are weaker than the large two-form background under the decoupling limit \cite{Ho:2013opa}. An action of the NS-NS D3-brane in the Poisson limit is  
\bea
{\cal L}_{NS-NS} \equiv\frac{1}{g_{YM}^2}\bigg(
-\frac{1}{4}{\cal F}_{\alpha\beta}^{\prime}{\cal F}^{\prime\alpha\beta}
-\frac{1}{2}{\cal F}_{\alpha\dot\mu}^{\prime}{\cal F}^{\prime\alpha\dot\mu}
-\frac{1}{4}{\cal F}_{\dot\mu\dot\nu}^{\prime}{\cal F}^{\prime\dot\mu\dot\nu}\bigg).
\label{action-NSNS}
\eea
We define the non-commutative field strength at the Poisson limit as
\bea
{\cal F}^{\prime}_{AB} 
\equiv
F^{\prime}_{AB}+g^{\prime}\{a^{\prime}_A, a^{\prime}_B\},
\eea
where $F^{\prime}_{AB} \equiv \partial_A a^{\prime}_B-\partial_B a^{\prime}_A$, $g^{\prime}\sim \theta^{\prime \dot1\dot2}$, $a^{\prime}_A$ can be identified as two-form gauge potential of the Nambu-Poisson M5-brane directly after we perform dimensional reduction, and the Poisson bracket is defined by
\be
\{f_1(x), f_2(x)\} \equiv 
\epsilon^{\dot\mu\dot\nu} \del_{\dot\mu}f_1 \del_{\dot\nu}f_2,
\ee
where $\eps^{\dot 1\dot2}=-\eps^{\dot2\dot1}=1$. Then we introduce an action of the R-R D3-brane in the Poisson limit as
\bea
\label{RR}
{\cal L}_{RR} = g_{YM}^2\bigg(-\frac{1}{2}{\cal H}^2_{\dot1\dot2}
+\frac{1}{2}{\cal F}_{\alpha\dot\mu}{\cal F}^{\alpha\dot\mu}
-\frac{1}{4}F_{\dot\mu\dot\nu}F^{\dot\mu\dot\nu}
+\frac{1}{2g}\epsilon^{\alpha\beta}{\cal F}_{\alpha\beta}\bigg),
\eea
where
\bea
{\cal H}_{\dot1\dot2}&\equiv&
H_{\dot1\dot2}+g\{b_{\dot1}, b_{\dot2}\},
\label{H-P} \\
{\cal F}_{\alpha\dot\mu}&\equiv&
(V^{-1})_{\dot\mu}{}^{\dot\nu}\bigg(F_{\alpha\dot\nu}
+gF_{\dot\nu\dot\sigma}\hat{B}_{\alpha}{}^{\dot\sigma}\bigg),
\label{Fadm-P} \\
{\cal F}_{\alpha\beta}&\equiv&
F_{\alpha\beta}+
g\bigg(
-F_{\alpha\dot\mu}\hat{B}_{\beta}{}^{\dot\mu}
-F_{\dot\mu\beta}\hat{B}_{\alpha}{}^{\dot\mu}
+gF_{\dot\mu\dot\nu}\hat{B}_{\alpha}{}^{\dot\mu}\hat{B}_{\beta}{}^{\dot\nu}
\bigg),
\label{Fab-P}
\eea
\bea
H_{\dot1\dot2} \equiv
\del_{\dot\mu}b^{\dot\mu},\qquad
V_{\dot\mu}{}^{\dot\nu}\equiv
\delta_{\dot\mu}{}^{\dot\nu}+g\eps^{\dot\nu\dot\lam}\partial_{\dot\mu}b_{\dot\lam},\qquad
F_{AB}\equiv
\partial_A a_B-\partial_B a_A,
\eea
and $\hat{B}_{\alpha}{}^{\dot\mu}$ satisfies
\bea
V_{\dot\mu}{}^{\dot\nu}
\bigg(\partial^{\alpha}b_{\dot\nu}-V^{\dot\rho}{}_{\dot\nu}\hat{B}^{\alpha}{}_{\dot\rho}\bigg)
+\epsilon^{\alpha\beta}F_{\beta\dot\mu}+g\epsilon^{\alpha\beta}F_{\dot\mu\dot\nu}
\hat{B}_{\beta}{}^{\dot\nu}=0.
\label{Bhat}
\eea
We will use a flat metric $\eta^{AB}$ to raise or lower indices. When we perform dimensional reduction in the Nambu-Poisson M5-brane theory to obtain the R-R D3-brane theory, we need to define  $\hat{B}_{\alpha}^{\dot\mu}$, which satisfies the non-linear equation \eqref{Bhat}, to identify $a_{\alpha}$. If we want to explicitly write  $\hat{B}_{\alpha}^{\dot\mu}$ in terms of other fields, we need to use perturbative method because it satisfies a nonlinear equation. However, we do not need to use perturbative method to find a field redefinition in the Poisson limit. Our field redefinition gives an exact equivalence between the NS-NS D3-brane and R-R D3-brane theories. 

To see the field redefinition between the NS-NS theory and the R-R theory in the Poisson limit, $b^{\dot\mu}$ in the R-R theory can be identified with $a^{\prime}_{\dot\mu}$ in the NS-NS theory according to
\be
g^2_{YM}b^{\dot\mu}\equiv\epsilon^{\dot\mu\dot\nu}a_{\dot\nu}^{\prime}.
\label{change-b-a}
\ee
Then we have
\bea
{\cal F}_{\dot1\dot2}^{\prime}
=F_{\dot1\dot2}^{\prime}+g^{\prime}\{ a_{\dot1}^{\prime}, a_{\dot2}^{\prime}\}
=g_{YM}^2{\cal H}_{\dot1\dot2},
\label{identify-F-H}
\eea
where we used $g^2_{YM}g^{\prime }=g$.
We can also find
\bea
\label{B1}
{\cal F}^{\prime}_{\a\dot\mu}=
-g_{YM}^2 \epsilon_{\dot\mu\dot\nu}\left(
\partial_{\alpha}b_{\dot\nu} 
- V_{\dot\lambda}{}^{\dot\nu}B_{\alpha}{}^{\dot\lambda}
\right),
\eea
where
\be
g_{YM}^2B_{\alpha}{}^{\dot\mu}\equiv
\epsilon^{\dot\mu\dot\nu}\partial_{\dot\nu}a'_{\alpha}.
\label{def-B}
\ee

Let us define
\bea
\label{B2}
{\cal F}^{\prime\prime}_{\alpha\dot\mu}\equiv
\epsilon_{\alpha\beta}{\cal F}^{\prime\beta\dot\nu}\epsilon_{\dot\nu\dot\mu}
= \epsilon_{\alpha\beta}\left(
\partial^{\beta}b_{\dot\mu} - B^{\beta\dot\nu}V_{\dot\nu\dot\mu}
\right),
\eea
which will be identified with ${\cal F}_{\alpha\dot\mu}$
in the R-R theory later after duality transformations.
After some change of variables, 
the Lagrangian (\ref{action-NSNS}) is equivalent to
\bea
\label{action3}
{\cal L}_{NS-NS} = 
-\frac{1}{4g_{YM}^2}{\cal F}_{\alpha\beta}^{\prime}{\cal F}^{\prime\alpha\beta}
+\frac{g_{YM}^2}{2}{\cal F}^{\prime\prime}_{\alpha\dot\mu}{\cal F}^{\prime\prime\alpha\dot\mu}
-\frac{g_{YM}^2}{2}{\cal H}_{\dot1\dot2}{\cal H}^{\dot1\dot2}.
\label{action-1}
\eea

The next step is to introduce an auxiliary field to dualize ${\cal F}_{\a\b}$.
Then the dynamics of the Lagrangian above is equivalent to the dynamics of the following Lagrangian
\bea
\label{action1}
{\cal L}_{NS-NS}^{(1)} \equiv
-\frac{g_{YM}^2}{2}\phi^2+\frac{1}{2}\epsilon^{\a\b}{\cal F}_{\a\b}^{\prime}\phi
+\frac{g_{YM}^2}{2}{\cal F}^{\prime\prime}_{\alpha\dot\mu}{\cal F}^{\prime\prime\alpha\dot\mu}
-\frac{g_{YM}^2}{2}{\cal H}_{\dot1\dot2}{\cal H}^{\dot1\dot2}.
\eea
An equation of motion of $\phi$ imposes a constraint 
\be
g_{YM}^2\phi = {\cal F}'_{01},
\ee
which replaces $\phi$ by ${\cal F}'_{01}$ so we go back to (\ref{action-1}).

Then we use field strength 
\be
F_{\dot\mu\dot\nu} \equiv \del_{\dot\mu}a_{\dot\nu} - \del_{\dot\mu}a_{\dot\nu}
\ee
to replace $\phi$.
We claim that the Lagrangian 
\bea
\label{action4}
{\cal L}^{(2)}_{NS-NS} = 
-\frac{g_{YM}^2}{2}F_{\dot1\dot2}^2
+\frac{1}{2}\epsilon^{\a\b}{\cal F}_{\a\b}^{\prime}F_{\dot1\dot2}
+\frac{g_{YM}^2}{2}{\cal F}^{\prime\prime}_{\alpha\dot\mu}{\cal F}^{\prime\prime\alpha\dot\mu}
-\frac{g_{YM}^2}{2}{\cal H}_{\dot1\dot2}{\cal H}^{\dot1\dot2}
\eea
is still equivalent to (\ref{action-1}).
An equation of motion of $a_{\dot\mu}$ implies
\be
\partial_{\dot\mu}\bigg(F_{\dot1\dot2}-{\cal F}_{01}^{\prime}\bigg)=0.
\label{eom-a}
\ee 
We assume that our fields vanish at infinities of the coordinates $x^{\dm}$, then we obtain
\bea
F_{\dot1\dot2}={\cal F}_{01}^{\prime}.
\eea

The last step is to carry out a duality transformation 
to get $a_{\alpha}$ from $a^{\prime}_{\a}$. 
Before that, we expand the second term in the Lagrangian (\ref{action4}) as
\bea
{\cal F}_{01}^{\prime}F_{\dot1\dot2}
&=&
F_{01}^{\prime}F_{\dot1\dot2} +g^{\prime} \{a_0^{\prime}, a_1^{\prime}\}F_{\dot1\dot2}
\nn \\
&=&g_{YM}^2\bigg( \epsilon^{\alpha\beta}\partial_{\beta}a_{\dot\mu}B_{\alpha}{}^{\dot\mu}
+ g\epsilon^{\alpha\beta}F_{\dot1\dot2}B_{\alpha}{}^{\dot1}B_{\beta}^{\dot2}
+ \mbox{total derivatives}\bigg).
\label{F01F12}
\eea
Then we replace $B_{\alpha}{}^{\dot\mu}$ by $\breve{B}_{\alpha}{}^{\dot\mu}$ 
in this Lagrangian and add an additional term
\bea
\label{additional term}
g_{YM}^2\epsilon^{\alpha\beta}f_{\beta\dot\mu}
\bigg( \breve{B}_{\alpha}{}^{\dot\mu}
-\epsilon^{\dot\mu\dot\nu}\partial_{\dot\nu}a'_{\alpha}
\bigg)
\eea
to perform a duality transformation.
The new Lagrangian is equivalent to the previous one (\ref{action4}) because
$\breve{B}_{\alpha}{}^{\dot\mu}$ equals to $B_{\alpha}{}^{\dot\mu}$ (\ref{def-B})
when the Lagrange multiplier $f_{\beta\dot\mu}$ is integrated out.
If we integrate $a^{\prime}_{\alpha}$ out, 
we obtain $\epsilon^{\dot\mu\dot\nu}\partial_{\dot\nu}f_{\beta\dot\mu}=0$. This implies that locally 
$f_{\beta\dot\mu} = - \partial_{\dot\mu}a_{\beta}$ 
for some field $a_{\beta}$. Therefore, (\ref{additional term}) becomes
\bea
-g_{YM}^2\epsilon^{\alpha\beta}\partial_{\dot\mu}a_{\beta}\breve{B}_{\alpha}{}^{\dot\mu}.
\eea
Now we integrate $\breve{B}_{\alpha}{}^{\dot\mu}$ out by the Gaussian integration. This result of the integration is equivalent to replacing  $\breve{B}_{\alpha}{}^{\dot\mu}$ by 
the solution of its equation of motion (\ref{Bhat}).

After integrating $\breve{B}_{\alpha}{}^{\dot\mu}$ out,
(\ref{F01F12}) can be rewritten as
\be
\frac{g_{YM}^2}{2g}\epsilon^{\alpha\beta}{\cal F}_{\alpha\beta}
\ee
up to total derivatives.
According to $ (\ref{B2})$, 
we find
\be
{\cal F}^{\prime\prime}_{\alpha\dot\mu}={\cal F}_{\alpha\dot\mu}.
\ee
Dynamics of the Lagrangian (\ref{action4}) is exactly equivalent to the dynamics of \eqref{RR}.
This Lagrangian can also be found from the Nambu-Poisson M5-brane theory by dimensional reduction. This redefinition should imply the meaning of the S-dualtiy or electric-magnetic duality.

\subsubsection{Perturbation}
We show an electric-magnetic duality from the R-R D3-brane theory to the NS-NS D3-brane theory up to the second order. We first mention how to perform the electric-magnetic duality up to the second order in a general Yang-Mills type theory by perturbation.
If the action is
\bea
g_{YM}^2\int d^4x~ \bigg(-\frac{1}{4}F_{AB}F^{AB}+gQ_1(F_{AB})+g^2Q_2(F_{AB})\bigg),\nn
\eea
we can consider 
\bea
\int d^4x~ \bigg(-\frac{g_{YM}^2}{4}F_{AB}F^{AB}+g_{YM}^2gQ_1(F_{AB})+g_{YM}^2g^2Q_2(F_{AB})+\frac{1}{2}\tilde{G}_{AB}F^{AB}\bigg)
\nn\\
\eea
to do the electric-magnetic duality. We add one additional term to promote $F_{AB}$ to an unconstrained field. Integrating $G=dB$ out, we obtain $dF=0$. Therefore, we solve $dF=0$ to obtain $F=dA$ to go back to the original theory. We vary $F_{AB}$ to obtain
\bea
-F^{AB}+g\frac{\delta Q_1}{\delta F_{AB}}+g^2\frac{\delta Q_2}{\delta F_{AB}}+\frac{1}{g_{YM}^2}\tilde{G}^{AB}=0\nn
\eea
or
\bea
F_{AB}=\frac{1}{g_{YM}^2}\tilde{G}_{AB}+g\frac{\delta Q_1}{\delta F^{AB}}+g^2\frac{\delta Q_2}{\delta F^{AB}}.\nn
\eea
Hence, we obtain an action of the form
\bea
&&\int d^4x~ \bigg\lbrack\frac{1}{4g_{YM}^2}\tilde{G}_{AB}\tilde{G}^{AB}+g_{YM}^2gQ_1\bigg(\frac{1}{g_{YM}^2}\tilde{G}_{AB}\bigg)+g_{YM}^2g^2Q_2\bigg(\frac{1}{g_{YM}^2}\tilde{G}_{AB}\bigg)
\nn\\
&&+\frac{g_{YM}^4g^2}{4}\frac{\delta Q_1}{\delta\tilde{G}_{AB}}\bigg(\frac{1}{g_{YM}^2}\tilde{G}_{AB}\bigg)
\frac{\delta Q_1}{\delta\tilde{G}^{AB}}\bigg(\frac{1}{g_{YM}^2}\tilde{G}_{AB}\bigg)\bigg\rbrack.
\eea
We used 
\bea
Q_1(F_{AB})\approx Q_1\bigg(\frac{1}{g_{YM}^2}\tilde{G}_{AB}\bigg)+\frac{g_{YM}^4g}{2}\frac{\delta Q_1}{\delta\tilde{G}_{AB}}\bigg(\frac{1}{g_{YM}^2}\tilde{G}_{AB}\bigg)
\frac{\delta Q_1}{\delta\tilde{G}^{AB}}\bigg(\frac{1}{g_{YM}^2}\tilde{G}_{AB}\bigg)\nn
\eea
in the above action.
We write the action in terms of $G_{AB}$ as
\bea
\int d^4x~\bigg( -\frac{1}{4g_{YM}^2}G_{AB}G^{AB}+g_{YM}^2gQ_1(G_{AB})+g_{YM}^2g^2Q_2(G_{AB})+
\frac{g_{YM}^4g^2}{4}\frac{\delta Q_1}{\delta\tilde{G}_{AB}}(G_{AB})\frac{\delta Q_1}{\delta\tilde{G}^{AB}}(G_{AB})
\bigg).\nn
\eea

We use the above electric-magnetic duality formula to go from the D3-brane in the large R-R two form background to the non-commutative $U(1)$ gauge theory or the D3-brane in the large NS-NS two-form background at the zeroth and first orders. This method relies on the fact that action does not contain gauge potential variables. For this goal, we fix
\bea
B^{\dot 1}=b^{\dot 2}=0.\nn
\eea

The action at the zeroth order is
\bea
\int d^4x~\bigg( -\frac{1}{4g_{YM}^2}G_{AB}G^{AB}\bigg)\nn
\eea
after we perform the electric-magnetic duality.
We integrate $b$ field out before we perform the electric-magnetic-duality. This result of integration is equivalent to setting
\bea
H_{\dot1\dot2}\approx -F_{01}.\nn
\eea

Then we consider first order correction. The action at the first order is given by
\bea
g_{YM}^2\int d^4x~\bigg(H_{\dot1\dot2}\epsilon^{\alpha\beta}F_{\alpha\dot1}\partial_{\beta}b^{\dot1}
-\frac{1}{2}\epsilon_{\alpha\beta}F_{\dot\mu\dot\nu}F^{\alpha\dot\mu}F^{\beta\dot\nu}
+F_{\alpha\dot\mu}F^{\alpha}{}_{\dot1}\partial^{\dot\mu}b^{\dot1}
-F_{\alpha\dot\mu}\partial^{\alpha}b_{\dot1}F^{\dot\mu\dot1}\bigg)\nn
\eea
up to total derivative terms.
We use
\bea
b^{\dot1}=\partial^{\dot1}\partial_{\dot1}^{-2}H_{\dot1\dot2}\nn
\eea
to replace $b^{\dot1}$ by $H_{\dot1\dot2}$.
Then we perform the electric-magnetic duality on this action at the first order as
\bea
&&\frac{1}{g_{YM}^2}\int d^4x~\bigg(\epsilon^{\alpha\beta}\tilde{G}_{01}\tilde{G}_{\alpha\dot 1}\partial_{\beta}\partial^{\dot 1}\partial_{\dot 1}^{-2}\tilde{G}_{01}-\frac{1}{2}\epsilon_{\alpha\beta}\tilde{G}_{\dot\mu\dot\nu}\tilde{G}^{\alpha\dot\mu}\tilde{G}^{\beta\dot\nu}
\nn\\
&&-\tilde{G}_{\alpha\dot\mu}\tilde{G}^{\alpha}{}_{\dot 1}\partial^{\dot\mu}\partial^{\dot 1}\partial_{\dot 1}^{-2}\tilde{G}_{01}+\tilde{G}_{\alpha\dot\mu}\tilde{G}^{\dot\mu\dot 1}\partial^{\alpha}\partial_{\dot 1}\partial_{\dot 1}{}^{-2}\tilde{G}_{01}\bigg).
\nn\\
\eea
For the purpose of rewriting a theory for $H_{\dot1\dot2}$, we have non-local operators (inverse derivatives) in our theory. But we know that the non-commutative $U(1)$ gauge theory at the Poisson limit should be described by local variables. Naively, this implies that we cannot obtain the non-commutative $U(1)$ gauge theory from this method. We will show that these non-local operators will be canceled.

The first term of the action at the first order is
\bea
&&\frac{1}{g_{YM}^2}\int d^4x~\bigg(-\epsilon^{\alpha\beta}\epsilon_{\alpha\gamma}G^{\dot1\dot2}
G^{\gamma\dot2}\partial_{\beta}\partial^{\dot 1}\partial_{\dot 1}^{-2}G^{\dot1\dot2}\bigg)=\frac{1}{g_{YM}^2}\int  d^4x~\bigg(-\epsilon^{\alpha\beta}\epsilon_{\alpha\gamma}\partial_{\dot1}B_{\dot2}
G^{\gamma\dot2}\partial_{\beta}\partial^{\dot 1}\partial_{\dot 1}^{-2}\partial_{\dot1}B_{\dot2}\bigg)
\nn\\
&=&\frac{1}{g_{YM}^2}\int d^4x~\bigg(G^{\beta\dot2}\partial_{\dot1}B_{\dot2}\partial_{\beta}B_{\dot2}\bigg).
\nn\\
\eea
The second tern of the action at the first order is 
\bea
&&\frac{1}{g_{YM}^2}\int d^4x~\bigg(-\epsilon_{\alpha\beta}\tilde{G}_{\dot1\dot2}\tilde{G}^{\alpha\dot1}
\tilde{G}^{\beta\dot2}\bigg)=\frac{1}{g_{YM}^2}\int d^4x~\bigg(\epsilon^{\beta\delta}G^{01}G^{\beta}{}_{\dot2}G_{\delta\dot1}\bigg)
\nn\\
&=&\frac{1}{g_{YM}^2}\int d^4x~\bigg(-\frac{1}{2}G^{\alpha\beta}\{B_{\alpha}, B_{\beta}\}-\epsilon^{\beta\delta}G^{01}\partial_{\dot 1}B_{\delta}\partial_{\beta}B_{\dot2}\bigg).
\eea
The third term of the action at the first order is
\bea
&&\frac{1}{g_{YM}^2}\int d^4x~\bigg(\epsilon_{\alpha\dot\mu\beta\dot\nu}\epsilon^{\alpha\dot1\gamma\dot2}
G^{\beta\dot\nu}G_{\gamma\dot2}
\partial^{\dot\mu}\partial^{\dot1}\partial_{\dot1}^{-2}G^{\dot1\dot2}\bigg)
=\frac{1}{g_{YM}^2}\int d^4x~\bigg(\epsilon_{\dot\mu\dot\nu}\epsilon_{\alpha\beta}\epsilon^{\alpha\gamma}G^{\beta\dot\nu}G_{\gamma\dot2}
\partial^{\dot\mu}\partial^{\dot1}\partial_{\dot1}^{-2}G^{\dot1\dot2}\bigg)
\nn\\
&=&\frac{1}{g_{YM}^2}\int d^4x~\bigg(-\partial^{\dot 1}B^{\gamma}G_{\gamma\dot2}\partial^{\dot2}B_{\dot2}-
G^{\gamma\dot2}G_{\gamma\dot2}\partial^{\dot1}B^{\dot2}\bigg).
\eea
The fourth term of the action at the first order is
\bea
&&\frac{1}{g_{YM}^2}\int d^4x~\bigg(\tilde{G}_{\alpha\dot2}\tilde{G}^{\dot2\dot1}\partial^{\alpha}\partial_{\dot1}
\partial_{\dot1}^{-2}\tilde{G}_{01}\bigg)
=\frac{1}{g_{YM}^2}\int d^4x~\bigg(-\tilde{G}_{\alpha\dot2}G^{01}\partial^{\alpha} B_{\dot2}\bigg)
\nn\\
&=&\frac{1}{g_{YM}^2}\int d^4x~\bigg(\epsilon_{\alpha\beta}G^{01}\partial_{\dot1}B^{\beta}\partial^{\alpha}B_{\dot2}\bigg).
\nn\\
\eea
We combine all four terms to obtain
\bea
\frac{1}{g_{YM}^2}\bigg(-G^{\alpha\dot2}\{B_{\alpha}, B_{\dot2}\}
-\frac{1}{2}G^{\alpha\beta}\{B_{\alpha}, B_{\beta}\}\bigg).\nn
\eea
This is the same as the D3-brane theory in the large NS-NS two-form background or non-commutative $U(1)$ gauge theory at the first order.

Now we consider the second order calculation of the electric-magnetic duality. This is a non-trivial consistent check due to the cancellation of the non-local operators . 

We first express an equation of motion of $H_{\dot 1\dot 2}$ up to first order,  
\bea
H_{\dot 1\dot 2}\approx-F_{01}+gA,\nn
\eea
where $A$ is a first order correction. Since $A$ contains many non-local operators, we use $A$ to denote the full terms instead of writing them explicitly. Now we want to obtain $Q_1(G_{AB})$ and $Q_2(G_{AB})$ from the zeroth and first orders action after we use the equation of motion of $H_{\dot 1\dot2}$. 
From the action at the zeroth order
\bea
g_{YM}^2\int d^4x~\bigg(-\frac{1}{2}H_{\dot 1\dot 2}H^{\dot 1\dot 2}-F_{01}H_{\dot1\dot 2}-\frac{1}{4}F_{\dot\mu\dot\nu}F^{\dot\mu\dot\nu}-\frac{1}{2}F_{\alpha\dot\mu}F^{\alpha\dot\mu}\bigg),\nn
\eea
we can obtain a term 
\bea
\frac{1}{g_{YM}^2}\int d^4x~\bigg(-\frac{1}{2}A^2\bigg),\nn
\eea
where we redefine $A$ by replacing $G$ by $\tilde{G}/g^2_{YM}$ in $Q_2(G_{AB})$.
Then the action at the first order 
\bea
g_{YM}^2\int d^4x~\bigg(\epsilon^{\alpha\beta}H_{\dot1\dot2}F_{\alpha\dot1}\partial_{\beta}b^{\dot1}
-\frac{1}{2}\epsilon_{\alpha\beta}F_{\dot\mu\dot\nu}F^{\alpha\dot\mu}F^{\beta\dot\nu}
+F_{\alpha\dot\mu}F^{\alpha}{}_{\dot1}\partial^{\dot\mu}b^{\dot1}
-F_{\alpha\dot\mu}\partial^{\alpha}b_{\dot1}F^{\dot\mu\dot1}\bigg)\nn
\eea
implies
\bea
Q_{1}&=&\frac{1}{g_{YM}^4}\bigg(
\epsilon^{\alpha\beta}\tilde{G}_{01}\tilde{G}_{\alpha\dot 1}\partial_{\beta}\partial^{\dot 1}\partial_{\dot 1}^{-2}\tilde{G}_{01}
-\epsilon_{\alpha\beta}\tilde{G}_{\dot1\dot2}\tilde{G}^{\alpha\dot1}\tilde{G}^{\beta\dot2}-\tilde{G}_{\alpha\dot1}\tilde{G}^{\alpha}{}_{\dot 1}\tilde{G}_{01}-\tilde{G}_{\alpha\dot2}\tilde{G}^{\alpha}{}_{\dot 1}\partial^{\dot2}\partial^{\dot 1}\partial_{\dot 1}^{-2}\tilde{G}_{01}\nn\\
&&-\tilde{G}_{\alpha\dot2}\tilde{G}^{\dot1\dot 2}\partial^{\alpha}\partial_{\dot 1}\partial_{\dot 1}{}^{-2}\tilde{G}_{01}\bigg),\nn\\
\eea
and one term for $Q_2(G_{AB})$ as
\bea
\frac{1}{g_{YM}^2}\int d^4x~\bigg(A^2\bigg),
\eea
where we also redefine $A$ by replacing $G$ by $\tilde{G}/g^2_{YM}$. Now we show that $A$ is canceled in our calculations.
\bea
\frac{\delta Q_1}{\delta \tilde{G}_{\dot 1\dot 2}}&=&\frac{1}{g_{YM}^4}\bigg(-\epsilon_{\alpha\beta}\tilde{G}^{\alpha\dot 1}\tilde{G}^{\beta\dot 2}-\tilde{G}_{\alpha\dot2}\partial^{\alpha}\partial_{\dot 1}\partial_{\dot 1}^{-2}\tilde{G}_{01}\bigg)=\frac{1}{g_{YM}^4}\bigg(\epsilon^{\alpha\beta}G_{\alpha\dot 1}G_{\beta\dot2}-\epsilon^{\alpha\beta}G_{\alpha\dot1}\partial_{\beta}B_{\dot2}\bigg)
\nn\\
&=&\frac{1}{g_{YM}^4}\bigg(\epsilon^{\alpha\beta}\partial_{\dot1}B_{\alpha}\partial_{\dot2}B_{\beta}\bigg),
\nn\\
\frac{\delta Q_1}{\delta\tilde{G}_{\alpha\dot 1}}&=&\frac{1}{g_{YM}^4}\bigg(\epsilon^{\alpha\beta}\tilde{G}_{01}\partial_{\beta}\partial^{\dot1}\partial_{\dot 1}^{-2}\tilde{G}_{01}-\epsilon^{\alpha\beta}\tilde{G}_{\dot1\dot2}\tilde{G}_{\beta\dot2}
-2\tilde{G}^{\alpha\dot1}\tilde{G}_{01}-\tilde{G}^{\alpha\dot2}\partial_{\dot2}\partial_{\dot1}
\partial_{\dot1}^{-2}\tilde{G}_{01}\bigg)
\nn\\
&=&\frac{1}{g_{YM}^4}\bigg(\epsilon^{\alpha\beta}G_{\dot1\dot2}\partial_{\beta}B_{\dot2}-G_{01}G^{\alpha\dot 1}-2\epsilon^{\alpha\beta}G_{\beta\dot2}G^{\dot1\dot2}
+\epsilon^{\alpha\beta}G_{\beta\dot1}\partial_{\dot2}B_{\dot2}\bigg),
\nn\\
\frac{\delta Q_1}{\delta\tilde{G}_{\alpha\dot 2}}&=&\frac{1}{g_{YM}^4}\bigg(\epsilon^{\alpha\beta}\tilde{G}_{\dot1\dot2}\tilde{G}_{\beta\dot1}-\tilde{G}^{\alpha\dot1}
\partial^{\dot2}\partial^{\dot1}\partial_{\dot1}^{-2}\tilde{G}_{01}
-\tilde{G}_{\dot1\dot2}\partial^{\alpha}\partial_{\dot1}\partial_{\dot1}^{-2}\tilde{G}_{01}\bigg)
\nn\\
&=&\frac{1}{g_{YM}^4}\bigg(-G_{01}G^{\alpha\dot2}-\epsilon^{\alpha\beta}G_{\beta\dot2}\partial_{\dot2}B_{\dot2}
+G_{01}\partial^{\alpha}B_{\dot2}\bigg)
\nn\\
&=&\frac{1}{g_{YM}^4}\bigg(G_{01}\partial_{\dot2}B^{\alpha}
-\epsilon^{\alpha\beta}G_{\beta\dot2}\partial_{\dot2}B_{\dot2}\bigg),
\nn\\
\frac{\delta Q_1}{\delta \tilde{G}_{01}}&=&\frac{1}{g_{YM}^4}A.
\eea
\bea
\frac{g_{YM}^4}{2}\frac{\delta Q_1}{\delta\tilde{G}_{\dot1\dot2}}\frac{\delta Q_1}{\delta\tilde{G}_{\dot1\dot2}}
&=&-\frac{1}{4g_{YM}^4}\{B^{\alpha}, B^{\beta}\}\{B_{\alpha}, B_{\beta}\},
\nn\\
\frac{g_{YM}^4}{2}\frac{\delta Q_1}{\delta\tilde{G}_{\alpha\dot1}}\frac{\delta Q_1}{\delta\tilde{G}^{\alpha\dot1}}
&=&\frac{1}{2g_{YM}^4}\bigg(-G_{\dot1\dot2}G_{\dot1\dot2}\partial_{\beta}B_{\dot2}\partial^{\beta}B_{\dot2}
+G_{01}G_{01}G^{\alpha\dot1}G_{\alpha\dot1}
-4G_{\beta\dot2}G^{\beta\dot2}G^{\dot1\dot2}G_{\dot1\dot2}
\nn\\
&&-G_{\beta\dot1}G^{\beta\dot1}\partial_{\dot2}B_{\dot2}\partial_{\dot2}B_{\dot2}
-2\epsilon^{\alpha\beta}G_{\dot1\dot2}G_{01}G_{\alpha\dot1}\partial_{\beta}B_{\dot2}
+4G_{\dot1\dot2}G_{\dot1\dot2}G^{\alpha\dot2}\partial_{\alpha}B_{\dot2}
\nn\\
&&-2G_{\dot1\dot2}G^{\alpha\dot1}\partial_{\alpha}B_{\dot2}\partial_{\dot2}B_{\dot2}
+4\epsilon^{\alpha\beta}G_{01}G_{\alpha\dot1}G_{\beta\dot2}G_{\dot1\dot2}
+4G_{\alpha\dot2}G_{\dot1\dot2}G^{\alpha\dot1}\partial_{\dot2}B_{\dot2}\bigg),
\nn\\
\frac{g_{YM}^4}{2}\frac{\delta Q_{1}}{\delta\tilde{G}_{\alpha\dot 2}}\frac{\delta Q_1}{\delta \tilde{G}^{\alpha\dot2}}
&=&\frac{1}{2g_{YM}^4}\bigg(G_{01}G_{01}\partial_{\dot2}B^{\alpha}\partial_{\dot2}B_{\alpha}
-G^{\alpha\dot2}G_{\alpha\dot2}\partial_{\dot2}B_{\dot2}\partial_{\dot2}B_{\dot2}
+2\epsilon^{\alpha\beta}G_{01}G_{\alpha\dot2}\partial_{\dot2}B_{\beta}\partial_{\dot2}B_{\dot2}\bigg),
\nn\\
\frac{g_{YM}^4}{2}\frac{\delta Q_{1}}{\delta\tilde{G}_{01}}\frac{\delta Q_1}{\delta \tilde{G}^{01}}
&=&-\frac{1}{2g_{YM}^4}A^2.
\eea
\bea
g_{YM}^2\bigg(\frac{1}{2g_{YM}^4}A^2+\frac{g_{YM}^4}{4}\frac{\delta Q_1}{\delta\tilde{G}_{AB}}\frac{\delta Q_1}{\delta\tilde{G}^{AB}}\bigg)&=&-\frac{1}{4g_{YM}^2}\{B^{\alpha}, B^{\beta}\}\{B_{\alpha}, B_{\beta}\}
\nn\\
&&+\frac{1}{2g_{YM}^2}\bigg(-G_{\dot1\dot2}G_{\dot1\dot2}\partial_{\beta}B_{\dot2}\partial^{\beta}B_{\dot2}
+G_{01}G_{01}G^{\alpha\dot1}G_{\alpha\dot1}
\nn\\
&&-4G_{\beta\dot2}G^{\beta\dot2}G^{\dot1\dot2}G_{\dot1\dot2}
-G_{\beta\dot1}G^{\beta\dot1}\partial_{\dot2}B_{\dot2}\partial_{\dot2}B_{\dot2}
-2\epsilon^{\alpha\beta}G_{\dot1\dot2}G_{01}G_{\alpha\dot1}\partial_{\beta}B_{\dot2}
\nn\\
&&+4G_{\dot1\dot2}G_{\dot1\dot2}G^{\alpha\dot2}\partial_{\alpha}B_{\dot2}
-2G_{\dot1\dot2}G^{\alpha\dot1}\partial_{\alpha}B_{\dot2}\partial_{\dot2}B_{\dot2}
\nn\\
&&+4\epsilon^{\alpha\beta}G_{01}G_{\alpha\dot1}G_{\beta\dot2}G_{\dot1\dot2}
+4G_{\alpha\dot2}G_{\dot1\dot2}G^{\alpha\dot1}\partial_{\dot2}B_{\dot2}
+G_{01}G_{01}\partial_{\dot2}B^{\alpha}\partial_{\dot2}B_{\alpha}
\nn\\
&&-G^{\alpha\dot2}G_{\alpha\dot2}\partial_{\dot2}B_{\dot2}\partial_{\dot2}B_{\dot2}
+2\epsilon^{\alpha\beta}G_{01}G_{\alpha\dot2}\partial_{\dot2}B_{\beta}\partial_{\dot2}B_{\dot2}\bigg).
\nn\\
\eea
After using the equation of motion for $H_{\dot1\dot2}$, we have non-local operators ($A$). Nevertheless, $A$ vanishes in the final result.

Let us start to calculate the action at the second order.
\bea
V_{\dot\mu}{}^{\dot\nu}&=&\delta_{\dot\mu}{}^{\dot\nu}+g\partial_{\dot\mu}b^{\dot\nu}, \qquad
(V_{\dot\mu}{}^{\dot\nu})^{-1}V_{\dot\nu}{}^{\dot\rho}=\delta_{\dot\mu}{}^{\dot\rho}, \qquad
(V_{\dot\mu}{}^{\dot\rho})^{-1}\approx\delta_{\dot\mu}{}^{\dot\rho}-g\partial_{\dot\mu}b^{\dot\rho}+g^2\partial_{\dot\mu}b^{\dot\nu}\partial_{\dot\nu}b^{\dot\rho}.
\nn\\
\eea
\bea
M_{\dot\mu\dot\nu}{}^{\alpha\beta}&=&V_{\dot\mu\dot\rho}V_{\dot\nu}{}^{\dot\rho}\delta^{\alpha\beta}-g\epsilon^{\alpha\beta}F_{\dot\mu\dot\nu}
\nn\\
&=&\delta_{\dot\mu\dot\nu}\delta^{\alpha\beta}+g\bigg( (\partial_{\dot\mu}b_{\dot\nu}
+\partial_{\dot\nu}b_{\dot\mu})\delta^{\alpha\beta}-\epsilon^{\alpha\beta}F_{\dot\mu\dot\nu}\bigg)
+g^2\partial_{\dot\mu}b_{\dot\rho}\partial_{\dot\nu}b^{\dot\rho}\delta^{\alpha\beta},
\nn\\
(M^{\dot\lambda\dot\mu}{}_{\gamma\alpha})^{-1}M_{\dot\mu\dot\nu}{}^{\alpha\beta}&=&
\delta^{\dot\lambda}{}_{\dot\nu}\delta_{\gamma}{}^{\beta},
\nn\\
(M^{\dot\lambda\dot\mu}{}_{\gamma\alpha})^{-1}&=&\delta^{\dot\lambda\dot\mu}\delta_{\gamma\alpha}-g\bigg\lbrack\bigg(\partial^{\dot\lambda}b^{\dot\mu}+\partial^{\dot\mu}b^{\dot\lambda}\bigg)\delta_{\gamma\alpha}
-\epsilon_{\gamma\alpha}F^{\dot\lambda\dot\mu}\bigg\rbrack
\nn\\
&&+g^2\Bigg\lbrack\bigg(\partial^{\dot\lambda}b^{\dot\rho}
\partial_{\dot\rho}b^{\dot\mu}+\partial^{\dot\rho}b^{\dot\lambda}\partial_{\dot\rho}b^{\dot\mu}
+\partial^{\dot\rho}b^{\dot\lambda}\partial^{\dot\mu}b_{\dot\rho}+F_{\dot\rho}{}^{\dot\mu}F^{\dot\lambda\dot\rho}\bigg)\delta_{\gamma\alpha}
\nn\\
&&-\bigg\lbrack\bigg(\partial^{\dot\lambda}b^{\dot\rho}+\partial^{\dot\rho}b^{\dot\lambda})F_{\dot\rho}{}^{\dot\mu}+(\partial_{\dot\rho}b^{\dot\mu}+\partial^{\dot\mu}b_{\dot\rho}\bigg)
F^{\dot\lambda\dot\rho}\bigg\rbrack\epsilon_{\gamma\alpha}\Bigg\rbrack.
\eea
\bea
\hat{B}_{\alpha}{}^{\dot\mu}&=&(M_{\alpha\beta}{}^{\dot\mu\dot\nu})^{-1}\bigg(V_{\dot\nu}{}^{\dot\sigma}\partial^{\beta}b_{\dot\sigma}+\epsilon^{\beta\gamma}F_{\gamma\dot\nu}\bigg),
\nn\\
V_{\dot\nu}{}^{\dot\sigma}\partial^{\beta}b_{\dot\sigma}+\epsilon^{\beta\gamma}F_{\gamma\dot\nu}
&=&\partial^{\beta}b_{\dot\nu}+\epsilon^{\beta\gamma}F_{\gamma\dot\nu}
+g\partial_{\dot\nu}b^{\dot\sigma}\partial^{\beta}b_{\dot\sigma},
\nn\\
\hat{B}_{\alpha}{}^{\dot\mu}&\approx&\partial_{\alpha}b^{\dot\mu}+\epsilon_{\alpha\beta}F^{\beta\dot\mu}
\nn\\
&&+g\bigg(-\partial^{\dot\nu}b^{\dot\mu}\partial_{\alpha}b_{\dot\nu}-\epsilon_{\alpha\beta}
\partial^{\dot\mu}b_{\dot\nu}F^{\beta\dot\nu}-\epsilon_{\alpha\beta}\partial_{\dot\nu}b^{\dot\mu}
F^{\beta\dot\nu}+\epsilon_{\alpha\beta}F^{\dot\mu\dot\nu}\partial^{\beta}b_{\dot\nu}
+F^{\dot\mu\dot\nu}F_{\alpha\dot\nu}\bigg)
\nn\\
&&+g^2\bigg(\partial^{\dot\rho}b^{\dot\mu}\partial^{\dot\nu}b_{\dot\rho}\partial_{\alpha}b_{\dot\nu}
+\epsilon_{\alpha\beta}F^{\beta\dot\nu}\partial^{\dot\mu}b^{\dot\rho}\partial_{\dot\rho}b_{\dot\nu}
+\epsilon_{\alpha\beta}F^{\beta\dot\nu}\partial^{\dot\rho}b^{\dot\mu}\partial_{\dot\rho}b_{\dot\nu}
+\epsilon_{\alpha\beta}F^{\beta\dot\nu}\partial^{\dot\rho}b^{\dot\mu}\partial_{\dot\nu}b_{\dot\rho}
\nn\\
&&-\epsilon_{\alpha\beta}F_{\dot\rho\dot\nu}\partial^{\dot\mu}b^{\dot\rho}\partial^{\beta}b^{\dot\nu}
-\epsilon_{\alpha\beta}F_{\dot\rho\dot\nu}\partial^{\dot\rho}b^{\dot\mu}\partial^{\beta}
b^{\dot\nu}
-\epsilon_{\alpha\beta}F^{\dot\mu\dot\rho}\partial^{\dot\nu}b_{\dot\rho}\partial^{\beta}b_{\dot\nu}
-F_{\alpha\dot\nu}F^{\dot\rho\dot\nu}\partial^{\dot\mu}b_{\dot\rho}
\nn\\
&&-F_{\alpha\dot\nu}F^{\dot\rho\dot\nu}\partial_{\dot\rho}b^{\dot\mu}
-F_{\alpha\dot\nu}F^{\dot\mu\dot\rho}\partial_{\dot\rho}b^{\dot\nu}
-F_{\alpha\dot\nu}F^{\dot\mu\dot\rho}\partial^{\dot\nu}b_{\dot\rho}
+F_{\dot\rho\dot\nu}F^{\dot\mu\dot\rho}\partial_{\alpha}b^{\dot\nu}
+\epsilon_{\alpha\beta}F^{\beta\dot\nu}F_{\dot\rho\dot\nu}F^{\dot\mu\dot\rho}
\bigg).
\nn\\
\eea
\bea
{\cal F}_{\alpha\dot\mu}&=&(V_{\dot\mu}{}^{\dot\nu})^{-1}\bigg(F_{\alpha\dot\nu}+gF_{\dot\nu\dot\delta}\hat{B}_{\alpha}{}^{\dot\delta}\bigg),
\nn\\
F_{\alpha\dot\nu}+gF_{\dot\nu\dot\delta}\hat{B}_{\alpha}{}^{\dot\delta}&=&
F_{\alpha\dot\nu}+g\bigg(F_{\dot\nu\dot\delta}\partial_{\alpha}b^{\dot\delta}
+\epsilon_{\alpha\beta}F_{\dot\nu\dot\delta}F^{\beta\dot\delta}\bigg)
\nn\\
&&+g^2\bigg(-F_{\dot\nu\dot\delta}\partial^{\dot\rho}b^{\dot\delta}\partial_{\alpha}b_{\dot\rho}
-\epsilon_{\alpha\beta}F_{\dot\nu\dot\delta}
\partial^{\dot\delta}b_{\dot\rho}F^{\beta\dot\rho}
-\epsilon_{\alpha\beta}F_{\dot\nu\dot\delta}\partial_{\dot\rho}
b^{\dot\delta}
F^{\beta\dot\rho}
+\epsilon_{\alpha\beta}F_{\dot\nu\dot\delta}F^{\dot\delta\dot\rho}\partial^{\beta}b_{\dot\rho}
\nn\\
&&+F_{\dot\nu\dot\delta}F^{\dot\delta\dot\rho}F_{\alpha\dot\rho}\bigg),
\nn\\
{\cal F}_{\alpha\dot\mu}&\approx&F_{\alpha\dot\mu}
+g\bigg(F_{\dot\mu\dot\delta}\partial_{\alpha}b^{\dot\delta}+\epsilon_{\alpha\beta} F_{\dot\mu\dot\delta}F^{\beta\dot\delta}-F_{\alpha\dot\nu}\partial_{\dot\mu}b^{\dot\nu}\bigg)
\nn\\
&&+g^2\bigg(-F_{\dot\mu\dot\delta}\partial^{\dot\rho}b^{\dot\delta}\partial_{\alpha}b_{\dot\rho}
-\epsilon_{\alpha\beta}F_{\dot\mu\dot\delta}\partial^{\dot\delta}b_{\dot\rho}F^{\beta\dot\rho}
-\epsilon_{\alpha\beta}F_{\dot\mu\dot\delta}\partial_{\dot\rho}b^{\dot\delta}F^{\beta\dot\rho}
+\epsilon_{\alpha\beta}F_{\dot\mu\dot\delta}F^{\dot\delta\dot\rho}\partial^{\beta}b_{\dot\rho}
\nn\\
&&+F_{\dot\mu\dot\delta}F^{\dot\delta\dot\rho}F_{\alpha\dot\rho}
-F_{\dot\nu\dot\delta}\partial_{\alpha}b^{\dot\delta}\partial_{\dot\mu}b^{\dot\nu}
-\epsilon_{\alpha\beta}F_{\dot\nu\dot\delta}F^{\beta\dot\delta}\partial_{\dot\mu}b^{\dot\nu}
+\partial_{\dot\mu}b^{\dot\rho}\partial_{\dot\rho}b^{\dot\nu}F_{\alpha\dot\nu}\bigg).
\eea
We calculate $g_{YM}^2\int d^4x~\bigg(\frac{1}{2}{\cal F}_{\alpha\dot\mu}{\cal F}^{\alpha\dot\mu}\bigg)$ at the second order.
\bea
&&g_{YM}^2\int d^4x~\bigg(\frac{1}{2}{\cal F}_{\alpha\dot\mu}{\cal F}^{\alpha\dot\mu}\bigg)
\nn\\
&\rightarrow&
g_{YM}^2\int d^4x~\bigg(-F_{\alpha\dot2}F^{\dot2\dot1}H_{\dot1\dot2}\big(\partial^{\alpha}
\partial_{\dot1}\partial_{\dot1}^{-2}
H_{\dot1\dot2}\big)
-3\epsilon^{\alpha\beta}F_{\alpha\dot2}F^{\dot2\dot1}F_{\beta\dot1}
H_{\dot1\dot2}
+2\epsilon^{\alpha\beta}F_{\alpha\dot1}F^{\dot1\dot2}F_{\dot2\dot1}
\big(\partial_{\beta}\partial_{\dot1}\partial_{\dot1}^{-2}
H_{\dot1\dot2}\big)
\nn\\
&&+\frac{3}{2}F_{\alpha\dot\mu}F^{\dot\mu\dot\delta}F_{\dot\delta\dot\rho}F^{\alpha\dot\rho}
-\epsilon^{\alpha\beta}F_{\alpha\dot1}F^{\dot1\dot2}F_{\beta\dot2}
H_{\dot1\dot2}+
\frac{3}{2}F_{\alpha\dot1}F^{\alpha\dot1}
H_{\dot1\dot2}H_{\dot1\dot2}
+F_{\alpha\dot2}F^{\alpha\dot1}\big(\partial^{\dot2}\partial_{\dot1}\partial_{\dot1}^{-2}
H_{\dot1\dot2}\big)H_{\dot1\dot2}
\nn\\
&&+\frac{1}{2}F_{\dot2\dot1}F^{\dot2\dot1}\big(\partial_{\alpha}\partial_{\dot1}\partial_{\dot1}^{-2}
H_{\dot1\dot2}\big)\big(\partial^{\alpha}
\partial_{\dot1}\partial_{\dot1}^{-2}
H_{\dot1\dot2}\big)
+\frac{1}{2}F_{\alpha\dot1}F^{\alpha\dot1}\big(\partial_{\dot2}\partial_{\dot1}\partial_{\dot1}^{-2}
H_{\dot1\dot2}\big)\big(\partial^{\dot2}\partial_{\dot1}\partial_{\dot1}^{-2}
H_{\dot1\dot2}\big)
\nn\\
&&-F_{\dot2\dot1}F^{\alpha\dot1}\big(\partial_{\alpha}\partial_{\dot1}\partial_{\dot1}^{-2}
H_{\dot1\dot2}\big)\big(\partial^{\dot2}\partial_{\dot1}\partial_{\dot1}^{-2}
H_{\dot1\dot2}\big)\bigg).
\nn
\eea
Then we calculate ${\cal F}_{\alpha\beta}$ at the third order because we have one term $\frac{1}{2g}\epsilon^{\alpha\beta}{\cal F}_{\alpha\beta}$ in our Lagrangian.
\bea
{\cal F}_{\alpha\beta}&=&F_{\alpha\beta}+g\bigg(-F_{\alpha\dot\mu}\hat{B}_{\beta}{}^{\dot\mu}-F_{\dot\mu\beta}\hat{B}_{\alpha}{}^{\dot\mu}\bigg)+g^2F_{\dot\mu\dot\nu}\hat{B}_{\alpha}{}^{\dot\mu}\hat{B}_{\beta}{}^{\dot\nu}.
\eea
We first calculate $-F_{\dot\mu\beta}\hat{B}_{\alpha}{}^{\dot\mu}$ at the second order.
\bea
&&-F_{\dot\mu\beta}\hat{B}_{\alpha}{}^{\dot\mu}
\nn\\
&\rightarrow&
-F_{\dot1\beta}\partial^{\dot1}b^{\dot1}\partial^{\dot1}b_{\dot1}\partial_{\alpha}
b_{\dot1}
-\epsilon_{\alpha\gamma}F_{\dot\mu\beta}F^{\gamma\dot1}\partial^{\dot\mu}b^{\dot1}
\partial_{\dot1}
b_{\dot1}
-\epsilon_{\alpha\gamma}F_{\dot1\beta}F^{\gamma\dot1}\partial^{\dot\rho}b^{\dot1}
\partial_{\dot\rho}
b_{\dot1}
\nn\\
&&-\epsilon_{\alpha\gamma}F_{\dot1\beta}F^{\gamma\dot\nu}\partial^{\dot1}b^{\dot1}
\partial_{\dot\nu}
b_{\dot1}
+\epsilon_{\alpha\gamma}F_{\dot1\beta}F_{\dot2\dot1}\partial^{\dot2}b^{\dot1}
\partial^{\gamma}
b^{\dot1}
+\epsilon_{\alpha\gamma}F_{\dot2\beta}F^{\dot2\dot1}\partial^{\dot1}b_{\dot1}
\partial^{\gamma}
b_{\dot1}
\nn\\
&&+F_{\dot1\beta}F_{\alpha\dot2}F^{\dot1\dot2}\partial^{\dot1}b_{\dot1}
+F_{\dot1\beta}F_{\alpha\dot\nu}F^{\dot\rho\dot\nu}\partial_{\dot\rho}b^{\dot1}
+F_{\dot\mu\beta}F_{\alpha\dot1}F^{\dot\mu\dot\rho}\partial_{\dot\rho}b^{\dot1}
+F_{\dot2\beta}F_{\alpha\dot\nu}F^{\dot2\dot1}\partial^{\dot\nu}b_{\dot1}
\nn\\
&&-F_{\dot1\beta}F_{\dot2\dot1}F^{\dot1\dot2}\partial_{\alpha}b^{\dot1}
-\epsilon_{\alpha\gamma}F_{\dot\mu\beta}F^{\gamma\dot\nu}F_{\dot\rho\dot\nu}
F^{\dot\mu\dot\rho}.
\eea
Then we calculate $F_{\dot\mu\dot\nu}\hat{B}_{\alpha}{}^{\dot\mu}\hat{B}_{\beta}{}^{\dot\nu}$ at the first order.
\bea
&&F_{\dot\mu\dot\nu}\hat{B}_{\alpha}{}^{\dot\mu}\hat{B}_{\beta}{}^{\dot\nu}
\nn\\
&\rightarrow&
\bigg(
-\epsilon_{\beta\rho}F_{\dot1\dot2}\partial_{\alpha}b^{\dot1}\partial^{\dot2}b_{\dot1}
F^{\rho\dot1}
+\epsilon_{\beta\rho}F_{\dot1\dot2}\partial_{\alpha}b^{\dot1}F^{\dot2\dot1}
\partial^{\rho}b_{\dot1}
+F_{\dot1\dot2}\partial_{\alpha}b^{\dot1}F^{\dot2\dot1}F_{\beta\dot1}
-\epsilon_{\alpha\gamma}F_{\dot2\dot1}F^{\gamma\dot2}\partial^{\dot1}b^{\dot1}\partial_{\beta}
b_{\dot1}
\nn\\
&&-
F_{\dot\mu\dot\nu}F_{\beta}{}^{\dot\mu}\partial^{\dot\nu}b^{\dot1}F_{\alpha\dot1}
-F_{\dot2\dot1}F_{\beta}{}^{\dot2}\partial_{\dot\rho}b^{\dot1}F_{\alpha}{}^{\dot\rho}
+
F_{\dot1\dot2}F_{\beta}{}^{\dot1}F^{\dot2\dot1}\partial_{\alpha}b_{\dot1}
+\epsilon_{\alpha\gamma}F_{\dot\mu\dot\nu}F^{\gamma\dot\mu}F^{\dot\nu\dot\rho}F_{\beta\dot\rho}
+
\epsilon_{\alpha\rho}F_{\dot1\dot2}\partial_{\beta}b^{\dot1}\partial^{\dot2}b_{\dot1}
F^{\rho\dot1}
\nn\\
&&-\epsilon_{\alpha\rho}F_{\dot1\dot2}\partial_{\beta}b^{\dot1}F^{\dot2\dot1}
\partial^{\rho}b_{\dot1}
-F_{\dot1\dot2}\partial_{\beta}b^{\dot1}F^{\dot2\dot1}F_{\alpha\dot1}
+\epsilon_{\beta\gamma}F_{\dot2\dot1}F^{\gamma\dot2}\partial^{\dot1}b^{\dot1}
\partial_{\alpha}
b_{\dot1}
+
F_{\dot\mu\dot\nu}F_{\alpha}{}^{\dot\mu}\partial^{\dot\nu}b^{\dot1}F_{\beta\dot1}
+F_{\dot2\dot1}F_{\alpha}{}^{\dot2}\partial_{\dot\rho}b^{\dot1}F_{\beta}{}^{\dot\rho}
\nn\\
&&-
F_{\dot1\dot2}F_{\alpha}{}^{\dot1}F^{\dot2\dot1}\partial_{\beta}b_{\dot1}
-\epsilon_{\beta\gamma}F_{\dot\mu\dot\nu}F^{\gamma\dot\mu}F^{\dot\nu\dot\rho}F_{\alpha\dot\rho}\bigg).
\eea
Hence, we obtain $g_{YM}^2\int d^4x~\bigg(\frac{1}{2g}\epsilon^{\alpha\beta}{\cal F}_{\alpha\beta}\bigg)$ at the second order.
\bea
&&g_{YM}^2\int d^4x~\bigg(\frac{1}{2g}\epsilon^{\alpha\beta}{\cal F}_{\alpha\beta}\bigg)
\nn\\
&\rightarrow&g_{YM}^2\int d^4x~
\bigg(-\epsilon^{\alpha\beta}F_{\dot1\beta}H_{\dot1\dot2}H_{\dot1\dot2}\big(\partial_{\alpha}
\partial_{\dot1}\partial_{\dot1}^{-2}H_{\dot1\dot2}\big)
+3F_{\dot1\beta}F^{\beta\dot1}H_{\dot1\dot2}
H_{\dot1\dot2}
+2F_{\dot2\beta}F^{\beta\dot1}\big(\partial^{\dot2}\partial_{\dot1}\partial_{\dot1}^{-2}H_{\dot1\dot2}\big)
H_{\dot1\dot2}
\nn\\
&&+
F_{\dot1\beta}F^{\beta\dot1}\big(\partial^{\dot2}\partial_{\dot1}\partial_{\dot1}^{-2}H_{\dot1\dot2}\big)
\big(\partial_{\dot2}
\partial_{\dot1}\partial_{\dot1}^{-2}H_{\dot1\dot2}\big)
-2F_{\dot1\beta}F_{\dot2\dot1}\big(\partial^{\dot2}\partial_{\dot1}\partial_{\dot1}^{-2}H_{\dot1\dot2}\big)
\big(\partial^{\beta}
\partial_{\dot1}\partial_{\dot1}^{-2}H_{\dot1\dot2}\big)
\nn\\
&&-2F_{\dot2\beta}F^{\dot2\dot1}H_{\dot1\dot2}
\big(\partial^{\beta}
\partial_{\dot1}\partial_{\dot1}^{-2}H_{\dot1\dot2}\big)
+6\epsilon^{\alpha\beta}F_{\dot1\beta}F_{\alpha\dot2}F^{\dot1\dot2}H_{\dot1\dot2}
+
F_{\dot1\dot2}\big(\partial_{\alpha}\partial_{\dot1}\partial_{\dot1}^{-2}H_{\dot1\dot2}\big)F^{\dot2\dot1}
\big(\partial^{\alpha}\partial_{\dot1}\partial_{\dot1}^{-2}H_{\dot1\dot2}\big)
\nn\\
&&+3\epsilon^{\alpha\beta}F_{\dot1\dot2}\big(\partial_{\alpha}\partial_{\dot1}\partial_{\dot1}^{-2}H_{\dot1\dot2}\big)
F^{\dot2\dot1}F_{\beta\dot1}
-2
F_{\dot\mu\dot\nu}F^{\beta\dot\mu}
F^{\dot\nu\dot\rho}F_{\beta\dot\rho}\bigg).
\eea
Let us show the term $g_{YM}^2\int d^4x~\bigg(\frac{1}{2}{\cal F}_{\alpha\dot\mu}{\cal F}^{\alpha\dot\mu}+\frac{1}{2g}\epsilon^{\alpha\beta}{\cal F}_{\alpha\beta}\bigg)$.
\bea
&&g_{YM}^2\int d^4x~\bigg(\frac{1}{2}{\cal F}_{\alpha\dot\mu}{\cal F}^{\alpha\dot\mu}+\frac{1}{2g}\epsilon^{\alpha\beta}{\cal F}_{\alpha\beta}\bigg)
\nn\\
&\rightarrow&g_{YM}^2\int d^4x~
\bigg(F_{\alpha\dot2}F^{\dot2\dot1}H_{\dot1\dot2}\big(\partial^{\alpha}
\partial_{\dot1}\partial_{\dot1}^{-2}
H_{\dot1\dot2}\big)
+3\epsilon^{\alpha\beta}F_{\alpha\dot2}F^{\dot2\dot1}F_{\beta\dot1}
H_{\dot1\dot2}
-
\epsilon^{\alpha\beta}F_{\alpha\dot1}F^{\dot1\dot2}F_{\dot2\dot1}\big(\partial_{\beta}\partial_{\dot1}\partial_{\dot1}^{-2}H_{\dot1\dot2}\big)
\nn\\
&&-
\frac{1}{2}F_{\alpha\dot\mu}F^{\dot\mu\dot\delta}F_{\dot\delta\dot\rho}F^{\alpha\dot\rho}
-\epsilon^{\alpha\beta}F_{\alpha\dot1}F^{\dot1\dot2}F_{\beta\dot2}
H_{\dot1\dot2}
-
\frac{3}{2}F_{\alpha\dot1}F^{\alpha\dot1}
H_{\dot1\dot2}H_{\dot1\dot2}
-F_{\alpha\dot2}F^{\alpha\dot1}\big(\partial^{\dot2}\partial_{\dot1}\partial_{\dot1}^{-2}
H_{\dot1\dot2}\big)H_{\dot1\dot2}
\nn\\
&&-\frac{1}{2}F_{\dot2\dot1}F^{\dot2\dot1}\big(\partial_{\alpha}\partial_{\dot1}\partial_{\dot1}^{-2}
H_{\dot1\dot2}\big)\big(\partial^{\alpha}
\partial_{\dot1}\partial_{\dot1}^{-2}
H_{\dot1\dot2}\big)
-\frac{1}{2}F_{\alpha\dot1}F^{\alpha\dot1}\big(\partial_{\dot2}\partial_{\dot1}\partial_{\dot1}^{-2}
H_{\dot1\dot2}\big)\big(\partial^{\dot2}\partial_{\dot1}\partial_{\dot1}^{-2}
H_{\dot1\dot2}\big)
\nn\\
&&+F_{\dot2\dot1}F^{\alpha\dot1}\big(\partial_{\alpha}\partial_{\dot1}\partial_{\dot1}^{-2}
H_{\dot1\dot2}\big)\big(\partial^{\dot2}\partial_{\dot1}\partial_{\dot1}^{-2}
H_{\dot1\dot2}\big)
-\epsilon^{\alpha\beta}F_{\dot1\beta}H_{\dot1\dot2}H_{\dot1\dot2}\big(\partial_{\alpha}
\partial_{\dot1}\partial_{\dot1}^{-2}H_{\dot1\dot2}\big)\bigg).
\eea
Then we use the equation of motion $H_{\dot1\dot2}\approx-F_{01}$ to express our action in terms of $F$.
\bea
&&g_{YM}^2\int d^4x~\bigg(F_{\alpha\dot2}F^{\dot2\dot1}F_{01}\big(\partial^{\alpha}
\partial_{\dot1}\partial_{\dot1}^{-2}
F_{01}\big)
-3\epsilon^{\alpha\beta}F_{\alpha\dot2}F^{\dot2\dot1}F_{\beta\dot1}
F_{01}
+
\epsilon^{\alpha\beta}F_{\alpha\dot1}F^{\dot1\dot2}F_{\dot2\dot1}\big(\partial_{\beta}\partial_{\dot1}\partial_{\dot1}^{-2}F_{01}\big)
\nn\\
&&-
\frac{1}{2}F_{\alpha\dot\mu}F^{\dot\mu\dot\delta}F_{\dot\delta\dot\rho}F^{\alpha\dot\rho}
+\epsilon^{\alpha\beta}F_{\alpha\dot1}F^{\dot1\dot2}F_{\beta\dot2}
F_{01}
-
\frac{3}{2}F_{\alpha\dot1}F^{\alpha\dot1}
F_{01}F_{01}
-F_{\alpha\dot2}F^{\alpha\dot1}\big(\partial^{\dot2}\partial_{\dot1}\partial_{\dot1}^{-2}
F_{01}\big)F_{01}
\nn\\
&&-\frac{1}{2}F_{\dot2\dot1}F^{\dot2\dot1}\big(\partial_{\alpha}\partial_{\dot1}\partial_{\dot1}^{-2}
F_{01}\big)\big(\partial^{\alpha}
\partial_{\dot1}\partial_{\dot1}^{-2}
F_{01}\big)
-\frac{1}{2}F_{\alpha\dot1}F^{\alpha\dot1}\big(\partial_{\dot2}\partial_{\dot1}\partial_{\dot1}^{-2}
F_{01}\big)\big(\partial^{\dot2}\partial_{\dot1}\partial_{\dot1}^{-2}
F_{01}\big)
\nn\\
&&+F_{\dot2\dot1}F^{\alpha\dot1}\big(\partial_{\alpha}\partial_{\dot1}\partial_{\dot1}^{-2}
F_{01}\big)\big(\partial^{\dot2}\partial_{\dot1}\partial_{\dot1}^{-2}
F_{01}\big)
+\epsilon^{\alpha\beta}F_{\dot1\beta}F_{01}F_{01}\big(\partial_{\alpha}
\partial_{\dot1}\partial_{\dot1}^{-2}F_{01}\big)\bigg).
\eea
Now we replace $F$ in terms of $\tilde{G}$.
\bea
&&\frac{1}{g_{YM}^2}\int d^4x~\bigg(\tilde{G}_{\alpha\dot2}\tilde{G}^{\dot2\dot1}\tilde{G}_{01}\big(\partial^{\alpha}
\partial_{\dot1}\partial_{\dot1}^{-2}
\tilde{G}_{01}\big)
-3\epsilon^{\alpha\beta}\tilde{G}_{\alpha\dot2}\tilde{G}^{\dot2\dot1}\tilde{G}_{\beta\dot1}
\tilde{G}_{01}
+
\epsilon^{\alpha\beta}\tilde{G}_{\alpha\dot1}\tilde{G}^{\dot1\dot2}\tilde{G}_{\dot2\dot1}\big(\partial_{\beta}\partial_{\dot1}\partial_{\dot1}^{-2}\tilde{G}_{01}\big)
\nn\\
&&-
\frac{1}{2}\tilde{G}_{\alpha\dot\mu}\tilde{G}^{\dot\mu\dot\delta}\tilde{G}_{\dot\delta\dot\rho}
\tilde{G}^{\alpha\dot\rho}
+\epsilon^{\alpha\beta}\tilde{G}_{\alpha\dot1}\tilde{G}^{\dot1\dot2}\tilde{G}_{\beta\dot2}
\tilde{G}_{01}
-
\frac{3}{2}\tilde{G}_{\alpha\dot1}\tilde{G}^{\alpha\dot1}
\tilde{G}_{01}\tilde{G}_{01}
-\tilde{G}_{\alpha\dot2}\tilde{G}^{\alpha\dot1}\big(\partial^{\dot2}\partial_{\dot1}\partial_{\dot1}^{-2}
\tilde{G}_{01}\big)\tilde{G}_{01}
\nn\\
&&-\frac{1}{2}\tilde{G}_{\dot2\dot1}\tilde{G}^{\dot2\dot1}\big(\partial_{\alpha}\partial_{\dot1}\partial_{\dot1}^{-2}
\tilde{G}_{01}\big)\big(\partial^{\alpha}
\partial_{\dot1}\partial_{\dot1}^{-2}
\tilde{G}_{01}\big)
-\frac{1}{2}\tilde{G}_{\alpha\dot1}\tilde{G}^{\alpha\dot1}\big(\partial_{\dot2}\partial_{\dot1}\partial_{\dot1}^{-2}
\tilde{G}_{01}\big)\big(\partial^{\dot2}\partial_{\dot1}\partial_{\dot1}^{-2}
\tilde{G}_{01}\big)
\nn\\
&&+\tilde{G}_{\dot2\dot1}\tilde{G}^{\alpha\dot1}\big(\partial_{\alpha}\partial_{\dot1}\partial_{\dot1}^{-2}
\tilde{G}_{01}\big)\big(\partial^{\dot2}\partial_{\dot1}\partial_{\dot1}^{-2}
\tilde{G}_{01}\big)
+\epsilon^{\alpha\beta}\tilde{G}_{\dot1\beta}\tilde{G}_{01}\tilde{G}_{01}\big(\partial_{\alpha}
\partial_{\dot1}\partial_{\dot1}^{-2}\tilde{G}_{01}\big)\bigg)
\nn\\
&=&\frac{1}{g_{YM}^2}\int d^4x~\bigg(-\epsilon_{\alpha\beta}G^{\beta\dot1}G_{01}G_{\dot1\dot2}\partial^{\alpha}
B_{\dot2}
-3\epsilon_{\alpha\beta}G^{\alpha\dot1}G_{01}G^{\beta\dot2}
G_{\dot1\dot2}
+
G^{\beta\dot2}G_{01}G_{01}\partial_{\beta}B_{\dot2}
\nn\\
&&-
\frac{1}{2}G_{01}G_{01}\partial_{\dot1}B^{\alpha}
\partial_{\dot1}B_{\alpha}
-\frac{1}{2}G_{01}G_{01}G^{\alpha\dot2}
G_{\alpha\dot2}
-\epsilon_{\alpha\beta}G^{\alpha\dot2}G_{01}G^{\beta\dot1}
G_{\dot1\dot2}
+
\frac{3}{2}G_{\alpha\dot2}G^{\alpha\dot2}
G_{\dot1\dot2}G_{\dot1\dot2}
\nn\\
&&-G_{\alpha\dot2}G^{\alpha\dot1}\partial^{\dot2}
B_{\dot2}G_{\dot1\dot2}
-\frac{1}{2}G_{01}G_{01}\partial_{\alpha}
B_{\dot2}\partial^{\alpha}
B_{\dot2}
+\frac{1}{2}G_{\alpha\dot2}G^{\alpha\dot2}\partial_{\dot2}
B_{\dot2}\partial_{\dot2}
B_{\dot2}
+\epsilon^{\alpha\beta}G_{01}G_{\beta\dot2}\partial_{\alpha}
B_{\dot2}\partial^{\dot2}
B_{\dot2}
\nn\\
&&-G^{\alpha\dot2}G_{\dot1\dot2}G_{\dot1\dot2}\partial_{\alpha}
B_{\dot2}\bigg).
\nn
\eea
Then we combine other terms.
\bea
&&\frac{1}{g_{YM}^2}\int d^4x~\bigg\lbrack-\epsilon_{\alpha\beta}G^{\beta\dot1}G_{01}G_{\dot1\dot2}\partial^{\alpha}
B_{\dot2}
-3\epsilon_{\alpha\beta}G^{\alpha\dot1}G_{01}G^{\beta\dot2}
G_{\dot1\dot2}
+G^{\alpha\dot2}G_{01}G_{01}\partial_{\alpha}B_{\dot2}
\nn\\
&&-
\frac{1}{2}G_{01}G_{01}\partial_{\dot1}B^{\alpha}
\partial_{\dot1}B_{\alpha}
-\frac{1}{2}G_{01}G_{01}G^{\alpha\dot2}
G_{\alpha\dot2}
-\epsilon_{\alpha\beta}G^{\alpha\dot2}G_{01}G^{\beta\dot1}
G_{\dot1\dot2}
+
\frac{3}{2}G_{\alpha\dot2}G^{\alpha\dot2}
G_{\dot1\dot2}G_{\dot1\dot2}
\nn\\
&&-G_{\alpha\dot2}G^{\alpha\dot1}\partial^{\dot2}
B_{\dot2}G_{\dot1\dot2}
-\frac{1}{2}G_{01}G_{01}\partial_{\alpha}
B_{\dot2}\partial^{\alpha}
B_{\dot2}
+\frac{1}{2}G_{\alpha\dot2}G^{\alpha\dot2}\partial_{\dot2}
B_{\dot2}\partial_{\dot2}
B_{\dot2}
+\epsilon^{\alpha\beta}G_{01}G_{\beta\dot2}\partial_{\alpha}
B_{\dot2}\partial^{\dot2}
B_{\dot2}
\nn\\
&&-G^{\alpha\dot2}G_{\dot1\dot2}G_{\dot1\dot2}\partial_{\alpha}
B_{\dot2}
-\frac{1}{4}\{B^{\alpha}, B^{\beta}\}\{B_{\alpha}, B_{\beta}\}
+\frac{1}{2}\bigg(-G_{\dot1\dot2}G_{\dot1\dot2}\partial_{\beta}B_{\dot2}\partial^{\beta}B_{\dot2}
+G_{01}G_{01}G^{\alpha\dot1}G_{\alpha\dot1}
\nn\\
&&-4G_{\beta\dot2}G^{\beta\dot2}G^{\dot1\dot2}G_{\dot1\dot2}
-G_{\beta\dot1}G^{\beta\dot1}\partial_{\dot2}B_{\dot2}\partial_{\dot2}B_{\dot2}
-2\epsilon^{\alpha\beta}G_{\dot1\dot2}G_{01}G_{\alpha\dot1}\partial_{\beta}B_{\dot2}
+4G_{\dot1\dot2}G_{\dot1\dot2}G^{\alpha\dot2}\partial_{\alpha}B_{\dot2}
\nn\\
&&
-2G_{\dot1\dot2}G^{\alpha\dot1}\partial_{\alpha}B_{\dot2}\partial_{\dot2}B_{\dot2}
+4\epsilon^{\alpha\beta}G_{01}G_{\alpha\dot1}G_{\beta\dot2}G_{\dot1\dot2}
+4G_{\alpha\dot2}G_{\dot1\dot2}G^{\alpha\dot1}\partial_{\dot2}B_{\dot2}G_{01}G_{01}\partial_{\dot2}
B^{\alpha}\partial_{\dot2}B_{\alpha}
\nn\\
&&
-G^{\alpha\dot2}G_{\alpha\dot2}\partial_{\dot2}B_{\dot2}\partial_{\dot2}B_{\dot2}
+2\epsilon^{\alpha\beta}G_{01}G_{\alpha\dot2}\partial_{\dot2}B_{\beta}\partial_{\dot2}B_{\dot2}\bigg)\bigg\rbrack
\nn\\
&=&\frac{1}{g_{YM}^2}
\int d^4x~\bigg(-\frac{1}{4}\{B^{\alpha}, B^{\beta}\}\{B_{\alpha}, B_{\beta}\}
-\frac{1}{2}\{B^{\alpha}, B^{\dot2}\}\{B_{\alpha}, B_{\dot2}\}\bigg)
.
\nn
\eea
To sum up, we obtain an expected answer at the second order. The calculations of the electric-magnetic duality at the second order use the equation of motion to replace $H$ by $F$. This is not equivalent to integrating out exactly. At the zeroth and first orders, we can perform the electric-magnetic duality exactly. At the second order, the electric-magnetic duality is a consistent check at classical level. Even for the classical consistent check, this is a non-trivial check for the equivalence between the R-R D3 and NS-NS D3 brane theories. The most difficult part is that the expansion for the R-R D3-brane theory up to the second order in this method. We eventually obtain the beautiful answer from the magical cancellation. The reason possibly comes from the covariant field strengths in the R-R D3-brane theory. Physical answer should only depend on on-shell degrees of freedom. It is why we have such a magical cancellation. This calculation is also interesting in the study of the non-local effects. If we do not employ any gauge fixing, we should find inverse derivative terms in our theory after performing the electric-magnetic duality. However, we find a consistent answer without the inverse derivative terms after gauge fixing. This implies that the non-local terms are not real physical non-local effects. These non-local effects just originated from gauge redundancy. We use gauge fixing to remove these inverse derivative terms. This result might have more physical implications in the gauge theory. The electric-magnetic duality is an equivalence between gauge coupling and inverse gauge coupling constants. This means that electric-magnetic duality is a non-perturbative duality. Our successful step is that we use $g$ to carry out the expansion. Small $g$ limit is equivalent to a large background limit. This expansion should avoid strong coupling problems. Although the non-commutative $U(1)$ gauge theory has a non-abelian-like structure, the non-commutative $U(1)$ gauge theory is still different from the non-abelian Yang-Mills theory. However, we can use the first method to perform the electric-magnetic duality on the non-abelian Yang-Mills and non-commutative $U(1)$ gauge theories. This shows that the first method should be a general way to perform the electric-magnetic duality. We will give more generic examples to perform the electric-magnetic duality by using the first method. The most interesting problem is to study the non-abelian gauge group in the third method. The motivation is a consistent construction of the multiple M5-branes theory. However, we encounter difficulties to apply this method to the non-abelian gauge group. When we perform the field redefinition in the non-commutative $U(1)$ gauge theory, we will dual a scalar field to a new field strength. We cannot use the same method in the non-abelian gauge group. This technical problem is similar with the Poincaré lemma in the non-abelian gauge theories. In our perturbation study, this method also encounters a similar problem. We believe that a consistent multiple M5-branes theory should have a totally different construction compared with the single M5-brane theory. The reason is due to the fact that the electric-magnetic duality in the non-abelian gauge group is different from the electric-magnetic duality in the abelian gauge group. The multiple M5-branes theory should have a consistent electric-magnetic duality in four dimensions after performing compactification on 2-torus. Thus, we believe that the multiple M5-branes theory possibly cannot be extended from the single M5-brane theory directly.

In the second method, we use the Seiberg-Witten map to rewrite the non-commutative $U(1)$ gauge theory from the commutative variables. Therefore, we obtain a similar form after we perform the electric-magnetic duality. In the third method, we always perform the electric-magnetic duality on the non-commutative space without using any commutative variables. Because they can be connected from the electric-magnetic duality or field redefinition, they should be equivalent theories after we perform the second and or third types electric-magnetic duality in the large background limit. Because the R-R D3-brane has a complicated action with the non-local inverse derivative operator. We should expect that we can use the perturbation method with respect to the non-commutativity parameter to find a non-local field redefinition to rewrite the R-R D3-brane with a compact form rewritten from the Poisson bracket. However, the non-local field redefinition is very hard to find systematically. When we use the perturbation to perform the electric-magnetic duality from the R-R D3-brane to the NS-NS D3-brane, we also use some techniques to remove non-local operators. If we consider the electric-magnetic duality from the NS-NS D3-brane to the R-R D3-brane, then the non-local operators will appear in our computation to bother us. Even if we know that it should work, the non-local operators have very difficult technique problems. In principle, we should determine their relations from perturbation methods at least up to the first order with respect to the non-commutativity parameter in the large background limit. We leave this interesting direction to the future. 

When we discuss the third method, we identify the NS-NS field with the R-R field. It is an interesting point because the second method needs to rewrite our theory in terms of the abelian field strength. Hence, the second method must be failed when you consider the non-abelian gauge theories. However, we find that the field redefinition and perturbation in the third method still cannot be extended to the non-abelian gauge theories for some steps because we need to dual a scalar field to a new field strength.

\section{Electric-Magnetic Duality in $p$-Form Gauge Theories and a Non-Commutative Theory with the Non-Abelian Structure}
\label{4}
In this section, we extend the first method of the electric-magnetic duality that we used in the non-commutative $U(1)$ gauge theory to the $p$-form theories and a non-commutative theory with the non-abelian structure. These studies should give a general extension to various types of simple theories.

\subsection{Abelian $p$-Form Theory}
The simplest abelian $p$-form theory is
\begin{equation}
S_{\mbox{ABp}}=-\frac{1}{2g_{YM}^2~(p+1)\,!}\int d^{2p+2} x~ F_{\mu_1\mu_2\cdots\mu_{p+1}} F^{\mu_1\mu_2\cdots\mu_{p+1}},
\end{equation}
where $F=dA$. We introduce an antisymmetric auxiliary field $G_{\mu_1\mu_2\cdots\mu_{p+1}}$. The action can be rewritten as
\begin{align}
 \frac{2}{(p+1)\,!}\int d^{2p+2} x~ \left(g_{YM}^2 G_{\mu_1\mu_2\cdots\mu_{p+1}} G^{\mu_1\mu_2\cdots\mu_{p+1}}-G^{\mu_1\mu_2\cdots\mu_{p+1}} F_{\mu_1\mu_2\cdots\mu_{p+1}}\right).
\end{align}
Then we integrate $A$ out to obtain
\bea
\frac{2}{(p+1)\,!}\int{\cal  D}G\exp\bigg\lbrack ig_{YM}^2\int d^{2p+2} x ~\bigg(G_{\mu_1\mu_2\cdots\mu_{p+1}} G^{\mu_1\mu_2\cdots\mu_{p+1}}\bigg)\bigg\rbrack\delta\bigg(\partial_{\nu_1}G^{\nu_1\nu_2
\cdots\nu_{p+1}}\bigg).
\nn\\
\eea
Solving the delta function is equivalent to finding
\bea
d\tilde{G}=0.
\eea
According to the Poincaré lemma, we get
\bea
\tilde{G}=d\tilde{A}.
\eea
Hence, we find
\bea
-\frac{g_{YM}^2}{2~(p+1)\,!}\int d^{2p+2} x~ \tilde{G}_{\mu_1\mu_2\cdots\mu_{p+1}} \tilde{G}^{\mu_1\mu_2\cdots\mu_{p+1}}.
\eea
Then we obtain
\bea
\partial_{\mu_1}F^{\mu_1\mu_2\cdots\mu_{p+1}}=0~\longleftrightarrow~\partial_{\mu_1}\tilde{G}^{\mu_1\mu_2\cdots\mu_{p+1}}=0
\eea
at classical level. Therefore, we generalize the electric-magnetic duality from one-form to $p$-form gauge potential in the abelian group. We can extend the electric-magnetic duality of the abelian $p$-form to all dimensions. Starting from
\bea
\frac{2}{(p+1)\,!}\int{\cal  D}G\exp\bigg\lbrack ig_{YM}^2\int d^{2p+2} x ~\bigg(G_{\mu_1\mu_2\cdots\mu_{p+1}} G^{\mu_1\mu_2\cdots\mu_{p+1}}\bigg)\bigg\rbrack\delta\bigg(\partial_{\nu_1}
G^{\nu_1\nu_2\cdots\nu_{p+1}}\bigg).
\nn\\
\eea
Introducing an auxiliary field $\tilde{A}$ to rewrite the partition function as
\bea
\int{\cal  D}G{\cal D}\tilde{A}\exp\bigg\lbrack i \frac{2g_{YM}^2}{(p+1)\,!}\int d^{2p+2} x ~\bigg(G_{\mu_1\mu_2\cdots\mu_{p+1}} G^{\mu_1\mu_2\cdots\mu_{p+1}}+(p+1)\tilde{A}_{\mu_2\mu_3\cdots\mu_{p+1}}\partial_{\mu_1}
G^{\mu_1\mu_2\cdots\mu_{p+1}}
\bigg)\bigg\rbrack,
\nn\\
\eea
where
$\tilde{A}$ is zero-form when $p=0$.
The last step is integrating $G$ out to get
\begin{equation}
-\frac{g_{YM}^2}{2~(p+1)\,!}\int d^{2p+2} x~ \tilde{G}^a_{\mu_1\mu_2\cdots\mu_{p+1}} \tilde{G}^{\mu_1\mu_2\cdots\mu_{p+1},a}.
\end{equation}
Because we do not use the Poincaré lemma to solve the delta function, we can extend the electric-magnetic duality to all dimensions for the abelian $p$-form theory in this method. This method can also be applied to the non-abelian $p$-form theory.

\subsection{Non-Abelian $p$-Form Theory}
The non-abelian $p$-form theory is
\begin{equation}
S_{\mbox{NABp}}=-\frac{1}{2g_{YM}^2~(p+1)\,!}\int d^{2p+2} x~ F^a_{\mu_1\mu_2\cdots\mu_{p+1}} F^{\mu_1\mu_2\cdots\mu_{p+1},a},
\end{equation}
where $F=DB$, $D\equiv d+A$, where $A$ is one-form gauge potential and $B$ is $p$-form gauge potential (If $p=1$, $B=A$). We introduce an antisymmetric auxiliary field, $G_{\mu_1\mu_2\cdots\mu_{p+1}}$ to rewrite the action as
\begin{align}
 \frac{2}{(p+1)\,!}\int d^{2p+2} x~ \left(g_{YM}^2G^a_{\mu_1\mu_2\cdots\mu_{p+1}} G^{\mu_1\mu_2\cdots\mu_{p+1},a}-G^{\mu_1\mu_2\cdots\mu_{p+1},a} F^a_{\mu_1\mu_2\cdots\mu_{p+1}}\right).
\end{align}
We integrate $A$ out to get
\bea
\int{\cal  D}G\ \exp\bigg\lbrack ig_{YM}^2 \frac{2}{(p+1)\,!}\int d^{2p+2} x ~\bigg(G^a_{\mu_1\mu_2\cdots\mu_{p+1}} G^{\mu_1\mu_2\cdots\mu_{p+1},a}\bigg)\bigg\rbrack\delta\bigg(D_{\nu_1}
G^{\nu_1\nu_2\cdots\nu_{p+1}}\bigg)
\nn\\
\eea
for $p\ne 1$.
Now we add one auxiliary field $\tilde{A}$ to rewrite the Lagrangian as
\bea
\int{\cal  D}G{\cal D}\tilde{A}\ \exp\bigg\lbrack i \frac{2g_{YM}^2}{(p+1)\,!}\int d^{2p+2} x ~\bigg(G^a_{\mu_1\mu_2\cdots\mu_{p+1}} G^{\mu_1\mu_2\cdots\mu_{p+1},a}+(p+1)\tilde{A}^a_{\mu_2\mu_3\cdots\mu_{p+1}}D_{\mu_1}
G^{\mu_1\mu_2\cdots\mu_{p+1}, a}
\bigg)\bigg\rbrack.
\nn
\eea
Then we integrate $G$ out to obtain the dual Lagrangian
\begin{equation}
-\frac{g_{YM}^2}{2~(p+1)\,!}\int d^{2p+2} x~ \tilde{G}^a_{\mu_1\mu_2\cdots\mu_{p+1}} \tilde{G}^{\mu_1\mu_2\cdots\mu_{p+1},a},
\end{equation}
where $\tilde{G}\equiv D\tilde{A}$. We obtain
\bea
D_{\mu_1}F^{\mu_1\mu_2\cdots\mu_{p+1},a}=0~\longleftrightarrow~ D_{\mu_1}\tilde{G}^{\mu_1\mu_2\cdots\mu_{p+1},a}=0
\eea
at classical level after we have performed the electric-magnetic duality. One can find that the electric-magnetic duality of the non-abelian one-form theory is more special than non-abelian higher-form theory. In the non-abelian one-form theory, the covariant derivative is also changed by the electric-magnetic duality, but the covariant derivative of the non-abelian higher-form theory does not. For a covariant property of the non-abelian higher form theory, we need to introduce an one-form gauge potential. This gauge potential is not affected by the electric-magnetic duality. But if one integrates this non-dynamical gauge potential out, this gauge potential should be related to the dynamical gauge potential. The dynamical potential should be affected by the electric-magnetic duality. We can explain that the electric-magnetic duality only duals the dynamical degrees of freedom in this method. Because we do not use the Poincaré lemma in the non-abelian $p$-form theory, this method can be applied to all dimensions in the non-abelian $p$-form theory although we denote dimensions to be $2p+2$ in our computations for each $p$.

\subsection{Non-Commutative Theory with the Non-Abelian Structure}
We start from
\begin{equation}
S_{\mbox{NCNA}}=-\frac{1}{4g_{YM}^2}\int d^4 x~ \hat{F}^a_{\mu\nu}*\hat{F}^{\mu\nu,a},
\end{equation}
where $\hat{F}^a_{\mu\nu}=\del_\mu \hat{A}^a_\nu-\del_\nu \hat{A}_\mu^a +\lbrack \hat{A}_\mu,\hat{A}_\nu\rbrack^a_*$, $\hat{A}_{\mu}\equiv\hat{A}_{\mu}^aT^a$, $T^a$ satisfies
\bea
T^aT^b-T^bT^a=f^{abc}T^c, \qquad T^aT^b+T^bT^a=d^{abc}T^c,
\eea
$\lbrack \hat{A}_{\mu}, \hat{A}_{\nu}\rbrack_*\equiv \lbrack \hat{A}_{\mu}, \hat{A}_{\nu}\rbrack^a_*T^a$,
and $*$ is the Moyal product. 

We rewrite our action by introducing an antisymmetric auxiliary field $\hat{G}^a_{\mn}$,  
\begin{equation}
\label{action}
S=\int d^4 x~ \left(g_{YM}^2\hat{G}^a_{\mn}*\hat{G}^{\mn,a}-\hat{G}^{\mn,a} *\hat{F}^a_{\mn}\right).
\end{equation}
We ignore the total derivative terms to express our action as
\begin{align}
S&=\int d^4 x~ \left( g_{YM}^2\hat{G}^a_{\mn}\hat{G}^{\mn,a}-\hat{G}^{\mn,a}\hat{F}^a_{\mn}\right) 
\nn\\
&=\int d^4 x ~\bigg\lbrack g_{YM}^2\hat{G}^a_{\mn}\hat{G}^{\mn,a}-\hat{G}^{\mn,a}\bigg(\del_\mu \hat{A}^a_\nu-\del_\nu \hat{A}^a_\mu +\lbrack\hat{A}_\mu,\hat{A}_\nu\rbrack^a_*\bigg)\bigg\rbrack\nn\\
&\approx \int d^4 x ~\Bigg\lbrack g_{YM}^2\hat{G}^a_{\mn}\hat{G}^{\mn,a}-\hat{G}^{\mn,a}\bigg\lbrack\del_\mu \hat{A}^a_\nu-\del_\nu \hat{A}^a_\mu +f^{abc}\hat{A}_{\mu}^b\hat{A}_{\nu}^c
+\frac{1}{2}\theta^{\rho\sigma}\bigg(\del_\rho \hat{A}_\mu \del_\sigma \hat{A}_\nu-\del_\rho \hat{A}_\nu \del_\sigma \hat{A}_\mu\bigg)^a\bigg\rbrack\Bigg\rbrack\nn\\
&=\int d^4 x ~\left(g_{YM}^2\hat{G}^a_{\mn}\hat{G}^{\mn,a}+2\hat{G}^{\mn,a}\del_\nu \hat{A}^a_\mu-f^{abc}\hat{G}^{\mn,a} \hat{A}_{\mu}^b\hat{A}_{\nu}^c
-d^{abc}\theta^{\rho\sigma}\hat{G}^{\mn,a} \del_\rho \hat{A}^b_\mu \del_\sigma \hat{A}^c_\nu\right)\nn\\
&=\int d^4 x ~\left(g_{YM}^2\hat{G}^a_{\mn}\hat{G}^{\mn,a}-2\del_\nu\hat{G}^{\mn,a} \hat{A}^a_\mu-f^{abc}\hat{G}^{\mn,a}\hat{A}_{\mu}^b\hat{A}_{\nu}^c
+d^{abc}\theta^{\rho\sigma} \hat{A}_\mu^b \del_\rho \hat{G}^{\mn,a}  \del_\sigma \hat{A}^c_\nu \right),
\end{align}
where we consider the Poisson limit for the Moyal product and ignore total derivative terms. We used the antisymmetric property of $\hat{G}^a_{\mn}$ and $\theta^{\rho\sigma}$, and integrate by part in our calculations. The action is quadratic in the field $\hat{A}$ so we can use the Gaussian integral \eqref{Gaussian} to integrate $\hat{A}$ out.

The partition function is given by
\bea
\label{Z1}
Z\sim \int \mathcal{D}G~ (\det M)^{-\frac{1}{2}} \exp \bigg\lbrack ig_{YM}^2\int d^4 x~\bigg( \hat{G}^a_{\mn}\hat{G}^{\mn,a}- \del_\gamma \hat{G}^{\mu\gamma,a} \big(M^{-1}\big)^{ab}_{\mn} \del_\lambda \hat{G}^{\nu\lambda,b} \bigg)\bigg\rbrack,
\nn\\
\eea
where $M^{\mn,bc}=-g_{YM}^2f^{abc}\hat{G}^{\mu\nu,a}+g_{YM}^2 d^{abc}\theta^{\rho\sigma}\del_\rho \hat{G}^{\mn,a}\del_\sigma\equiv -f^{\prime abc}\hat{G}^{\mu\nu,a}+d^{abc}\tilde{\theta}^{\rho\sigma}\del_\rho \hat{G}^{\mn,a}\del_\sigma$. We use $\bar{A}^a_\mu\equiv(M^{-1})^{ab}_{\mn} \del_\rho \hat{G}^{\nu\rho,b}$ to let $\hat{G}^{\mn,a}$ satisfies the equation of motion in the Poisson limit as
\begin{align}
&\del_\nu \hat{G}^{\nu\mu,a}+M^{\mn,ab}\bar{A}^b_\nu=0 \nn\\
\Rightarrow& \del_\nu \hat{G}^{\nu\mu,a}-f^{\prime abc}G^{\mu\nu,c}\bar{A}_{\nu}^b+d^{abc}\tilde{\theta}^{\rho\sigma}\del_\rho \hat{G}^{\mn,c}\del_\sigma \bar{A}^b_\nu=0 \nn\\
\Rightarrow& \del_\nu \hat{G}^{\nu\mu}+\lbrack \bar{A}_{\nu}, \hat{G}^{\nu\mu}\rbrack
+ \{\bar{A}_\nu,\hat{G}^{\nu\mu}\}=0,
\end{align}
where $\bar{A}_{\nu}\equiv T^a\bar{A}_{\nu}^a$ and $\hat{G}^{\nu\mu}\equiv T^a\hat{G}^{\nu\mu,a}$.

Then we ignore total derivative term to rewrite the action as
\begin{align*}
\int d^4x~\hat{G}^{\mn,a}\hat{F}^a_{\mn}(\bar{A})&\approx g_{YM}^2\int d^4x~\hat{G}^{\mn,a}\bigg(\del_\mu \bar{A}^a_\nu-\del_\nu \bar{A}^a_\mu + f^{\prime abc}\bar{A}_{\mu}^b\bar{A}_{\nu}^c
+ \{\bar{A}_\mu,\bar{A}_\nu\}^a\bigg)\nn\\
&=\int d^4x~\bigg(-2 \hat{G}^{\mn,a}\del_\nu \bar{A}^a_\mu +f^{\prime abc}\hat{G}^{\mu\nu,a}\bar{A}_{\mu}^b\bar{A}_{\nu}^c
+d^{abc}\hat{G}^{\mn,a} \tilde{\theta}^{\rho\sigma}\del_\rho \bar{A}^b_\mu \del_\sigma \bar{A}^c_\nu\bigg) \nn\\
&=\int d^4x~\bigg(2\del_\nu \hat{G}^{\mn,a} \bar{A}^a_\mu +f^{\prime abc}\hat{G}^{\mu\nu,a}\bar{A}_{\mu}^b\bar{A}_{\nu}^c
-d^{abc}\bar{A}^b_\mu \tilde{\theta}^{\rho\sigma} \del_\rho \hat{G}^{\mn,a}\del_\sigma \bar{A}^c_\nu\bigg) \nn\\
&=\int d^4x~\bigg(2 \del_\nu \hat{G}^{\mn,a} \big(M^{-1}\big)^{ab}_{\mu\lambda} \del_\rho \hat{G}^{\lambda\rho,b}-\big(M^{-1}\big)^{be}_{\mu\lambda} \del_\rho \hat{G}^{\lambda\rho,e} (M)^{\mn,bc}\big(M^{-1}\big)^{cd}_{\nu\sigma}\del_\delta \hat{G}^{\sigma\delta,d}\bigg) \nn\\
&=\int d^4x~\del_\nu \hat{G}^{\mn,a} (M^{-1})^{ab}_{\mu\lambda} \del_\rho \hat{G}^{\lambda\rho,b},
\end{align*}
where we used integration by part, and $\bar{A}^a_\mu=(M^{-1})^{ab}_{\mn} \del_\rho \hat{G}^{\nu\rho,b}$. This term is equal to the second term in \eqref{Z1}.
Therefore, we obtain alternative form of the partition function as
\bea
\label{Z2}
Z\sim\int \mathcal{D}G ~(\det M)^{-\frac{1}{2}} \int \mathcal{D}\bar{A}~\exp \left( ig_{YM}^2\int d^4 x~\Big( \hat{G}^a_{\mu\nu}\hat{G}^{\mu\nu,a}- \hat{G}^{\mu\nu,a}\hat{F}_{\mu\nu}^a(\bar{A}) \Big) \right)
\nn\\
\delta\bigg(2\bar{A}^a_\rho-2(M^{-1})^{ab}_{\rho\sigma} \del_\lambda \hat{G}^{\sigma\lambda,b}\bigg),
\eea
where the factor of 2 does not affect the calculation. The delta function can be expressed as
\begin{equation}
\delta(2\bar{A}-2M^{-1}\del \hat{G})=\delta\Big(M^{-1}(2M\bar{A}-2\del \hat{G})\Big),
\end{equation}
where $\bar{A}\equiv \bar{A}_{\mu}^a$, $M^{-1}\equiv \big(M^{-1}\big)_{\mu\nu}^{ab}$ and $\partial\hat{G}\equiv \partial_{\lambda}\hat{G}^{\mu\lambda,a}$. We used the matrix notation to simplify our index notations. This extracts an additional factor $\det M$ out of the delta function after integrating . Hence, we obtain
\bea
\label{Z3}
Z\sim \int \mathcal{D}G ~(\det M)^{\frac{1}{2}} \int \mathcal{D}\bar{A}\mathcal{D}\Lambda~\exp \Bigg\lbrack ig_{YM}^2\int d^4 x~\bigg\lbrack \hat{G}^a_{\mu\nu}\hat{G}^{\mu\nu,a}- \hat{G}^{\mu\nu,a}\hat{F}^a_{\mu\nu}(\bar{A})
\nn\\
-\Lambda^a_\mu \bigg(2M^{\mu\nu,ab}\bar{A}^b_\nu-2\del_\rho \hat{G}^{\mu\rho,a}\bigg) \bigg\rbrack \Bigg\rbrack.
\nn\\
\eea
Now we simplify the term in the last bracket as
\bea
&&\int d^4x~2\bigg\lbrack\Lambda^a_\mu\bigg(M^{\mn,ab}\bar{A}^b_\nu-\del_\rho\hat{G}^{\mu\rho,a}\bigg)\bigg\rbrack
\nn\\
&=&\int d^4x~\bigg(2\Lambda^a_\mu M^{\mn,ab}\bar{A}^b_\nu+2\del_\rho \Lambda^a_\mu\hat{G}^{\mu\rho,a}\bigg)
\nn\\
&=&\int d^4x~\bigg(2f^{\prime abc}\Lambda_{\mu}^a\bar{A}_{\nu}^bG^{\mu\nu,c}+2d^{abc}\Lambda^a_\mu \tilde{\theta}^{\rho\sigma}\del_\rho \hat{G}^{\mn,c}\del_\sigma \bar{A}^b_\nu+2\del_\rho \Lambda^a_\mu\hat{G}^{\mu\rho,a}\bigg) 
\nn\\
&=&\int d^4x~\bigg(2f^{\prime abc}G^{\mu\nu,a}\bar{A}_{\mu}^b\Lambda_{\nu}^c- 2d^{abc}\hat{G}^{\mn,c} \tilde{\theta}^{\rho\sigma} \del_\rho \Lambda^a_\mu \del_\sigma \bar{A}^b_\nu+2\del_\rho \Lambda^a_\mu\hat{G}^{\mu\rho,a}\bigg)
\nn\\
&=&\int d^4x~\bigg\lbrack2\hat{G}^{\mn,a} \times\bigg(\lbrack\bar{A}_{\nu}, \Lambda_{\mu}\rbrack^a+ \{\bar{A}_\nu,\Lambda_\mu\}^a\bigg) +2(\del_\nu \Lambda^a_\mu)\hat{G}^{\mu\nu,a}\bigg\rbrack
\nn\\
&=&\int d^4x~\bigg(2\hat{G}^{\mn,a} \big(D_\nu^{(\bar{A})} \Lambda_\mu\big)^a\bigg) 
\nn\\
&=&\int d^4x~\bigg(-2\hat{G}^{\mn,a} \big(D_\mu^{(\bar{A})} \Lambda_\nu\big)^a\bigg),
\eea
where we define $D_\mu^{(\bar{A})}O\equiv\del_\mu O+\lbrack\bar{A}_{\mu}, O\rbrack+\{\bar{A}_\mu, O\}$ and $\Lambda_\mu\equiv\Lambda_{\mu}^aT^a$. Substitution of this term into the partition function gives
\begin{equation}
Z\approx\int \mathcal{D}G ~(\det M)^{\frac{1}{2}} \int \mathcal{D}\bar{A}\mathcal{D}\Lambda~\exp \Bigg\lbrack ig_{YM}^2\int d^4 x~\bigg\lbrack \hat{G}^{\mn,a}\bigg(\hat{G}^a_{\mn}- \hat{F}^a_{\mn}(\bar{A})+2\big(D_\mu^{(\bar{A})} \Lambda_\nu\big)^a \bigg) \bigg\rbrack \Bigg\rbrack.
\end{equation}
Let us define a variable $\tilde{A}_\mu\equiv\bar{A}_\mu-\Lambda_\mu$. The field strength can be written as
\begin{align}
\hat{F}_{\mn}(\bar{A})\approx&\del_\mu(\tilde{A}_\nu+\Lambda_\nu)-\del_\nu(\tilde{A}_\mu+\Lambda_\mu)+\lbrack \tilde{A}_{\mu}+\Lambda_{\mu}, \tilde{A}_{\nu}+\Lambda_{\nu}\rbrack
+\{\tilde{A}_\mu+\Lambda_\mu,\tilde{A}_\nu+\Lambda_\nu\}
\nn\\
=&\del_\mu \tilde{A}_\nu-\del_\nu \tilde{A}_\mu+\lbrack \tilde{A}_{\mu}, \tilde{A}_{\nu}\rbrack+\{\tilde{A}_\mu,\tilde{A}_\nu\}
\nn\\
&+\del_\mu \Lambda_\nu-\del_\nu \Lambda_\mu +\lbrack\tilde{A}_{\mu}, \Lambda_{\nu}\rbrack+\lbrack\Lambda_{\mu}, \tilde{A}_{\nu}\rbrack+\lbrack\Lambda_{\mu}, \Lambda_{\nu}\rbrack
+\{\tilde{A}_\mu,\Lambda_\nu\}+\{\Lambda_\mu,\tilde{A}_\nu\}+\{\Lambda_\mu,\Lambda_\nu\} \nn\\
=&\hat{F}_{\mn}(\tilde{A})+D_\mu^{(\bar{A})} \Lambda_\nu-D_\nu^{(\bar{A})} \Lambda_\mu +\lbrack\Lambda_{\mu}, \Lambda_{\nu}\rbrack+\{\Lambda_{\mu}, \Lambda_{\nu}\},
\end{align}
where $\hat{F}\equiv\hat{F}^aT^a$.
Thus, we obtain
\begin{align}
Z\approx &\int \mathcal{D}G ~(\det M)^{\frac{1}{2}} \int \mathcal{D}\tilde{A}\mathcal{D}\Lambda~\exp \Bigg\lbrack ig_{YM}^2\int d^4 x~\bigg\lbrack \hat{G}^{\mn,a}\bigg(\hat{G}^a_{\mn}- \hat{F}^a_{\mn}\big(\tilde{A}\big)-\lbrack \Lambda_{\mu}, \Lambda_{\nu}\rbrack^a
-\{\Lambda_{\mu}, \Lambda_{\nu}\}^a\bigg) \bigg\rbrack \Bigg\rbrack\nn\\
=&\int \mathcal{D}G ~(\det M)^{\frac{1}{2}} \int \mathcal{D}\tilde{A}~\exp \Bigg\lbrack ig_{YM}^2\int d^4 x~\bigg\lbrack \hat{G}^{\mn,a}\bigg(\hat{G}^a_{\mn}- \hat{F}^a_{\mn}\big(\tilde{A}\big)\bigg)\bigg\rbrack \Bigg\rbrack 
\nn\\
&\times\int \mathcal{D}\Lambda ~\exp \Bigg\lbrack -ig_{YM}^2\int d^4x~\bigg\lbrack\hat{G}^{\mn,a}\bigg(\lbrack\Lambda_{\mu}, \Lambda_{\nu}\rbrack^a+\{\Lambda_{\mu}, \Lambda_{\nu}\}^a\bigg)\bigg\rbrack \Bigg\rbrack\nn\\
=&\int \mathcal{D}G ~(\det M)^{\frac{1}{2}} \int \mathcal{D}\tilde{A}~\exp \Bigg\lbrack ig_{YM}^2\int d^4 x~\bigg\lbrack \hat{G}^{\mn,a}\bigg(\hat{G}^a_{\mn}- \hat{F}^a_{\mn}\big(\tilde{A}\big)\bigg) \bigg\rbrack\Bigg\rbrack
\nn\\
&\times\int \mathcal{D}\Lambda ~\exp \bigg\lbrack ig_{YM}^2\int d^4x~ \bigg(d^{abc}\Lambda^b_\mu \tilde{\theta}^{\rho\sigma}\del_\rho \hat{G}^{\mn,a}  \del_\sigma \Lambda^c_\nu-f^{\prime abc}\Lambda_{\mu}^b\hat{G}^{\mu\nu,a}\Lambda^c_{\nu}\bigg) \bigg\rbrack\nn\\
=&\int \mathcal{D}G ~(\det M)^{\frac{1}{2}} \int \mathcal{D}\tilde{A}~\exp \Bigg\lbrack ig_{YM}^2\int d^4 x~\bigg\lbrack \hat{G}^{\mn,a}\bigg(\hat{G}^a_{\mn}- \hat{F}^a_{\mn}\big(\tilde{A}\big)\bigg)\bigg\rbrack \Bigg\rbrack 
\nn\\
&\times\int \mathcal{D}\Lambda ~\exp \bigg(i g_{YM}^2\int d^4x~ \Lambda^a_\mu M^{\mn,ab} \Lambda_\nu^b \bigg)\nn\\
\sim &\int \mathcal{D}G \mathcal{D}\tilde{A}~\exp \Bigg\lbrack ig_{YM}^2\int d^4 x~\bigg\lbrack \hat{G}^{\mn}\bigg(\hat{G}_{\mn}- \hat{F}_{\mn}\big(\tilde{A}\big)\bigg)\bigg\rbrack \Bigg\rbrack .
\end{align}
We eventually integrate the field $\Lambda$ out and obtain a factor $(\det M)^{-1/2}$ to cancel the factor $(\det M)^{1/2}$ in front of the measure. This calculation shows 
\bea
D_{\mu}^{(A)}\hat{F}^{\mu\nu}(A)=0~\longleftrightarrow~D_{\mu}^{\tilde{(A)}}\hat{F}^{\mu\nu}(\tilde{A})=0
\eea
at classical level. This method does not use the Poincaré lemma, we can extend from four dimensions to all dimensions. Although the non-abelian structure is different from the non-commutative structure, we can use the first method of the electric-magnetic dualities in the non-commutative $U(1)$ gauge theory to define the electric-magnetic duality for this kind of theory. Other methods cannot be applied to this theory. The second method of the electric-magnetic dualities in the non-commutative $U(1)$ gauge theory relies on the Seiberg-Witten map. If a theory has a non-abelian structure, then this theory should have degrees of freedom on gauge potentials. When we perform the field redefinition to relate two theories for the non-commutative $U(1)$ gauge theory in the third method, the field redefinition is related to the gauge potentials. From this point of view, we can use this method to perform the electric-magnetic duality for a theory with the non-abelian structure. Unfortunately, this method still relies on a dual. This dual is valid for the ordinary derivative. When considering the covariant derivative, this dual cannot be used. For non-abelian gauge theories, we do not know how to define a covariant field strength by using the ordinary derivative. The third method naively uses a field redefinition, which is related to the gauge potentials, to perform the electric-magnetic duality, but it still relies on some properties that exist only in the abelian gauge theories. This study shows that the non-abelian gauge theories have a more delicate structure than the non-commutative structures in the electric-magnetic dualities. 

The most interesting aspect in the electric-magnetic dualities should be the multiple M5-branes theory. A low energy effective theory of the multiple M5-branes does not have a suitable or totally consistent Lagrangian formulation. If we compactify two torus with different ordering, we obtain two multiple D3-branes theories. Two multiple D3-branes theories should be related to each other via the electric-magnetic duality or S-duality. A consistent electric-magnetic duality should motivate us to find the multiple M5-branes theory. In our studies, we use some ways to find a suitable or workable definition. We should define an electric-magnetic duality related to gauge potentials, but this is not enough. We also need to understand how to dual a scalar field to field strength in non-abelian gauge theories. These difficulties should also appear in the construction of the multiple M5-branes theory. From our results, we find that the electric-magnetic dualities of non-abelian gauge theories should be totally different from the electric-magnetic dualities of abelian gauge theories. This points out the form of the multiple M5-branes should be very different from the single M5-brane theory. If we perform the electric-magnetic duality by the first method, it should not be hard to find the consistency for the Lagrangian formulation between the NS-NS multiple D3-branes and the R-R multiple D3-branes. Based on the T-duality, we can find the R-R multiple D$p$-branes for the Lagrangian formulation. They should be easy to construct. The most difficult thing is how to find the multiple M5-branes such that we can get the multiple D3-branes in the NS-NS or R-R backgrounds by compactification. The problem comes from the dualization for the non-abelian gauge theories. This problem also occurs in the electric-magnetic dualities. A study of the electric-magnetic duality reveals the main problem for the Lagrangian formulation of the multiple M5-branes. We leave the further studies in the multiple M5-branes to future works.

\section{Discussion and Conclusion}
\label{5}
We study the electric-magnetic dualities in gauge theories by using path integration. The electric-magnetic duality for the abelian Yang-Mills theory can be understood as exchanging electric and magnetic fields in path integration like the Maxwell's equations. We define the electric-magnetic duality for the abelian Yang-Mills theory by
\bea
\partial_{\mu}F^{\mu\nu}=0~\longleftrightarrow ~\partial_{\mu}\tilde{F}^{\mu\nu}=0, \qquad dF=0~\rightarrow F_{\mu\nu}=\frac{1}{2}\epsilon^{\mu\nu\rho\sigma}\tilde{F}_{\rho\sigma}.
\eea
The first relation is an invariant equation of motion under the electric-magnetic duality and the second relation relates the field strength to the dual field strength in four dimensions by using the Poincaré lemma. Especially for the second relation, this is a strong condition to restrict dimensionality for the electric-magnetic dualities in path integration formulation. If we want to define the electric-magnetic dualities without using the second relation, we can extend the electric-magnetic dualities from four dimensions to all dimensions. Naively, this is still a suitable definition for the electric-magnetic dualities. One should think about the degrees of freedom between the electric and magnetic fields. In the abelian one-form Yang-Mills theory, we have an equal number of degrees of freedom in both the electric and magnetic fields in four dimensions. This means that the electric-magnetic dualities lose the standard meaning in other dimensions (other than four dimensions). In four dimensions, we have possibilities to find a map between the electric (or magnetic) field and dual magnetic (or electric) field. But we do not have this kind of map in other dimensions. If we want to maintain the standard meaning of the electric-magnetic dualities, the Poincaré lemma should be important. If we only replace $d$ by $D=d+A$ in the Poincaré lemma, this lemma should not be valid without putting in more conditions. Then a direct generalization from the abelian Yang-Mills theory to the non-abelian Yang-Mills theory should be impossible. A definition or operation must have workable or calculable properties. Before we give a clever  definition, we use a workable or calculable definition without using too restricted conditions. In other words, we only use the first condition to define the electric-magnetic dualities in non-abelian gauge theories. This might not be a smart definition to define the electric-magnetic dualities in non-abelian gauge theories, but this should be calculable. A smart definition should have a restriction on dimensionality without losing the standard meaning of the electric-magnetic dualities. However, we do not have this kind of lemma at non-abelian level. Even without this lemma, the electric-magnetic dualities still exchange strong and weak coupling constants for the non-abelian gauge theories in this method. We can map the ordinary gauge theories to the dual gauge theories by exchanging the ordinary and dual gauge fields, and using ordinary electric and magnetic fields simultaneously to find the dual electric or magnetic fields. A main problem in the non-abelian gauge theories comes from the covariant property. In the abelian gauge theories, the equations of motion do not depend on gauge potentials, but the non-abelian gauge theories do. This is why we lose the Poincaré lemma in the non-abelian gauge theories. Dependence on gauge potentials  implies that exchanging the electric and magnetic fields is not a suitable operation for the electric-magnetic dualities. But this does not mean that we cannot have a modified Poincaré lemma to put restrictions on dimensionality. We believe that the electric-magnetic dualities should work in four dimensions with equal degrees of freedom between the electric and magnetic fields for the non-abelian Yang-Mills theory. The non-abelian $p$-form theory has one interesting feature in the electric-magnetic duality. In order to have a gauge covariant property, we need to introduce a non-dynamical gauge potential except for the one-form gauge potential. Then we find that the electric-magnetic duality does not dual the non-dynamical degrees of freedom. Since electric-magnetic dualities have different physical meanings for different methods, we perform three methods on the non-commutative $U(1)$ gauge theory and compare their different physical implications. The non-commutative $U(1)$ gauge theory has a non-abelian-like structure which comes from the Moyal product and this theory can be described by the field strength without using gauge potentials. The non-commutative $U(1)$ gauge theory simultaneously has two interesting properties so we can compare meanings in different electric-magnetic dualities. In the first method, we do not have restrictions on dimensionality, but we have the same form of action after performing the electric-magnetic duality. The ordinary electric and magnetic fields, and dual electric and magnetic fields are covariant quantities. From a symmetry point of view, electric and magnetic fields being covariant field strength should be nice. In the second method, we use the Seiberg-Witten map to rewrite our theory in terms of abelian field strength. This symmetry structure helps us to avoid difficulties of the non-abelian-like structure. Due to this rewriting, we have restrictions on the number of dimensions. In the third method, we consider large background limit in the non-commutative theories. We use field redefinition and perturbation to study the electric-magnetic duality. If one naively performs the electric-magnetic duality, one will find non-locality in the dual action. However, this non-locality should not be real because we can use a suitable gauge fixing to remove them. We perform the exact calculation up to the first order. At the second order, our calculations only concern the classical information. The electric-magnetic dualities did not extend to this order due to the non-Gaussian effects being difficult to handle. However, we obtain a consistent result and give a string interpretation to this duality. The electric-magnetic dualities invert the coupling constant so we cannot use the perturbation method to study the electric-magnetic duality. The primary reason is due to the fact that our perturbative parameter is the non-commutativity parameter (large antisymmetric background). Even if we go into the strongly coupled regime under electric-magnetic duality, the dual effective theory is still a well-defined theory under the decoupling limit. In this comparison, one should find that the non-commutative $U(1)$ gauge theory is different from the non-abelian gauge theories although they have the similar structure. Due to this reason, we also perform the electric-magnetic dualities in the non-commutative theory with the non-abelian structure. The first method we used in the non-commutative $U(1)$ gauge theory is still applicable in this kind of model. This kind of theory should have some applications in the multiple branes theory. This is also our motivation to study the non-commutative theory with the non-abelian structure. However, our studies should provide a generic analysis for electric-magnetic dualities in path integral formulation.

One important problem related to the electric-magnetic dualities is the multiple M5-branes. One consistency check in the multiple M5-branes theory is on the multiple D3-branes for the electric-magnetic dualities after compactifying 2-torus with different orderings. A low energy effective theory of the multiple D3-branes on the non-commutative space should be the non-commutative Yang-Mills theory at leading order. If we believe that the first method we used in the non-commutative $U(1)$ gauge theory is a good definition for the electric-magnetic dualities, we already obtained the consistency for the electric-magnetic dualities. One problem in the multiple D$p$-branes is the effective action in a large R-R background limit. So far we did not have a consistent action based on gauge symmetry, T-duality and S-duality in the Poisson limit. Based on these conditions, this model should not be difficult to construct. The main non-trivial consistency is an expected duality between two-form gauge potential in the multiple M5-branes and one-form gauge potential in the multiple D4-branes. We leave this interesting work to the future. 

The Nambu-Poisson M5-brane provides a R-R D3 brane from dimensional reduction. Because the Nambu-Poisson M5-brane is valid at the second order, the R-R D3-brane cannot go beyond this order. A conjecture for the full order is given, but symmetry (gauge symmetry and supersymmetry) is not totally understood to all orders. A complete study should give a complete action. This should give us a motivation to check the electric-magnetic duality for the R-R D3-brane to all orders. This study should motivate many low-energy effective theories in many different aspects.

The most important and fundamental issue is how to improve definition of the electric-magnetic dualities for the non-abelian gauge theories. In abelian gauge theories, we relate electric (magnetic) fields to dual magnetic (electric) fields in path integral formulation. From an equation of motion in the non-abelian Yang-Mills theory, exchanging electric and magnetic fields should not be a suitable operation for the electric-magnetic dualities in the non-abelian Yang-Mills theory. When treating an one-form gauge potential in the electric-magnetic duality, one has a non-trivial determinant factor in partition function. Based on this non-trivial factor, a modified Poincaré lemma is difficult to define in path integral formulation. We have a no-go theorem \cite{Deser:1976iy} to show that the electric-magnetic duality cannot be performed on the non-abelian Yang-Mills theory with an invariant equation of motion and the Poincaré lemma. However, a condition of restricting dimensionality allows us to keep the standard meaning for electric-magnetic dualities. The other approach is to modify the definitions of the electric and magnetic fields in the non-abelian gauge theories. A quantity which can be observed should be gauge invariant. We do not insist on a gauge covariant definition for field strength. A main problem in the electric-magnetic dualities of non-abelian gauge theories comes from the gauge covariant property. The gauge covariant property also lead to the ambiguities of the entanglement entropy. The entanglement entropy \cite{Casini:2011kv} in gauge theories is not a gauge invariant quantity in a tensor product decomposition of the Hilbert space. A proposal is to consider non-tensor product decomposition with a non-trivial center between two regions. In the abelian gauge theories, this proposal should be well-understood. For the non-abelian gauge theories, the entanglement entropy may suffer from the gauge covariant problem. Defining a gauge invariant entanglement entropy will be difficult. This direction should help us understand more about holograph, black hole, and thermal entropy \cite{Myers:1986un}. Candidates of gauge invariant quantities are $\det({F})$ and Wilson loop. A full gauge invariant construction should be interesting and could affect our understanding of gauge theories from different ways. Another approach of electric-magnetic duality is to include all spin fields with general relativity \cite{Deser:2014ssa}. In this case, we do not use duality rotation to perform the electric-magnetic duality.

\section*{Acknowledgement}

We would like to thank Dah-Wei Chiou, S. Deser and Xing Huang for their useful discussion.

\end{document}